\documentclass[aps, twocolumn, prd, longbibliography, showkeys]{revtex4-2}

\usepackage{amsmath}
\usepackage{amssymb}
\usepackage{braket}
\usepackage{bm}
\usepackage{cuted}
\usepackage{dcolumn}
\usepackage{diagbox}
\usepackage{epsfig}
\usepackage{esint}
\usepackage{flushend}
\usepackage{graphicx}
\usepackage{mathrsfs}
\usepackage{mathtools}
\usepackage{newtxmath}
\usepackage{newtxtext}
\usepackage{slashed}
\usepackage{subfigure}
\usepackage[pagewise, mathlines]{lineno}     
\usepackage[pagebackref=false, colorlinks=true]{hyperref}
\hypersetup{
linkcolor=red,     
citecolor=blue,     
filecolor=blue,        
urlcolor=magenta}      

\newcommand{\hammer}{{\textsc{h-amr}}}

\newcommand{\ta}{{\texttt{T0}}}
\newcommand{\tb}{{\texttt{T15}}}
\newcommand{\tc}{{\texttt{T30}}}
\newcommand{\td}{{\texttt{T45}}}
\newcommand{\te}{{\texttt{T60}}}
\newcommand{\tf}{{\texttt{T75}}}
\newcommand{\tg}{{\texttt{T90}}}
\newcommand{\ti}{{\texttt{T90H}}}
\newcommand{\tj}{{\texttt{T90L}}}
\newcommand{\tiltBHa}{{\texttt{T0BH}}}
\newcommand{\tiltBHb}{{\texttt{T60BH}}}
\newcommand{\tiltBHc}{{\texttt{T90BH}}}
\newcommand{\tiltBHd}{{\texttt{T90BHb}}}
\newcommand{\rg}{{r_{\rm g}}}
\newcommand{\mdot}{{\dot{M}}}

\def\araa{ARA\&A}                
\def\apj{ApJ}                    
\def\apjl{ApJL}                  
\def\apjs{ApJS}                  

\def\aap{A\&A}                   
\def\mnras{MNRAS}                
\def\nat{Nature} 

\def\pasj{PASJ}                  

\def\ssr{Space Science Reviews}
\def\physrep{Physics Reports}
\def\nar{New Astronomy Reviews}
\def\na{New Astronomy}

\begin{document}

\preprint{APS/123-QED}

\title{Misaligned magnetized accretion flows onto spinning black holes:\\ Magneto-spin alignment, outflow power and intermittent jets}

\author{Koushik Chatterjee$^{1,2,3}$}
\email{kchatt@umd.edu}
\author{Nicholas Kaaz$^{4}$}
\email{nkaaz@u.northwestern.edu}
\author{Matthew Liska$^{5,6}$}
\email{mliska3@gatech.edu}
\author{Alexander Tchekhovskoy$^{4}$}
\email{atchekho@northwestern.edu}
\author{Sera Markoff$^{7,8}$}
\email{S.B.Markoff@uva.nl}

\affiliation{
$^1$Department of Physics, University of Maryland, 7901 Regents Drive, College Park, Maryland 20742, USA\\
$^2$Black Hole Initiative at Harvard University, 20 Garden Street, Cambridge, Massachusetts 02138, USA\\
$^3$Harvard-Smithsonian Center for Astrophysics, 60 Garden Street, Cambridge, Massachusetts 02138, USA\\
$^4$Center for Interdisciplinary Exploration \& Research in Astrophysics (CIERA), Physics \& Astronomy, Northwestern University, Evanston, Illinois 60202, USA\\
$^5$Center for Relativistic Astrophysics, Georgia Institute of Technology, Howey Physics Building, 837 State Street NorthWest, Atlanta, Georgia 30332, USA\\
$^6$Institute for Theory and Computation, Harvard University, 60 Garden Street, Cambridge, Massachusetts 02138, USA\\
$^7$Anton Pannekoek Institute for Astronomy, University of Amsterdam, Science Park 904, 1098 XH Amsterdam, The Netherlands\\
$^8$Gravitation Astroparticle Physics Amsterdam (GRAPPA) Institute, University of Amsterdam, Science Park 904, 1098 XH Amsterdam, The Netherlands
}

\date{\today}

\begin{abstract}
Magnetic fields regulate black hole (BH) accretion, governing both inflow and outflow dynamics. When a BH accumulates substantial vertical magnetic flux, it enters the magnetically arrested disk (MAD) state, where dynamically important fields power jets and trigger disk eruptions. We investigate MAD evolution when the BH spin and disk angular momentum are misaligned, a likely scenario in many BH systems. Using numerical simulations, we show that jets from rapidly spinning, prograde BHs realign the inner disk via the magneto-spin alignment mechanism for initial tilts up to $\mathcal{T} \lesssim 60^\circ$. Larger tilts lead to intermittent jets that disrupt the disk out to $r\gtrsim100$ gravitational radii, creating hot cavities and magnetized filaments. These episodic jets form a mini$-$feedback loop and may explain quasiperiodic X-ray and radio flares observed in low-luminosity active galaxies. We also find that (i) BH spin and disk tilt influence the amount of magnetic flux accumulated at the horizon, and (ii) large-scale, thick, misaligned accretion flows do not exhibit sustained Lense$-$Thirring (LT) precession. This suggests that slowly accreting BHs ($\dot{M} \ll 10^{-3} \dot{M}_{\rm Edd}$) are unlikely to show lightcurve quasiperiodic oscillations from LT precession, consistent with observations. Instead, magnetic flux eruptions drive jet wobbling and lateral motion, offering an alternative explanation for phenomena such as the M87 jet’s apparent precession and rapid swings in blazar jet orientation.
\end{abstract}

\maketitle


\section{\label{sec:intro}Introduction} 

\begin{figure*}
    \centering
    \includegraphics[width=\textwidth,trim= 3in 1in 2.3in 0in, clip]{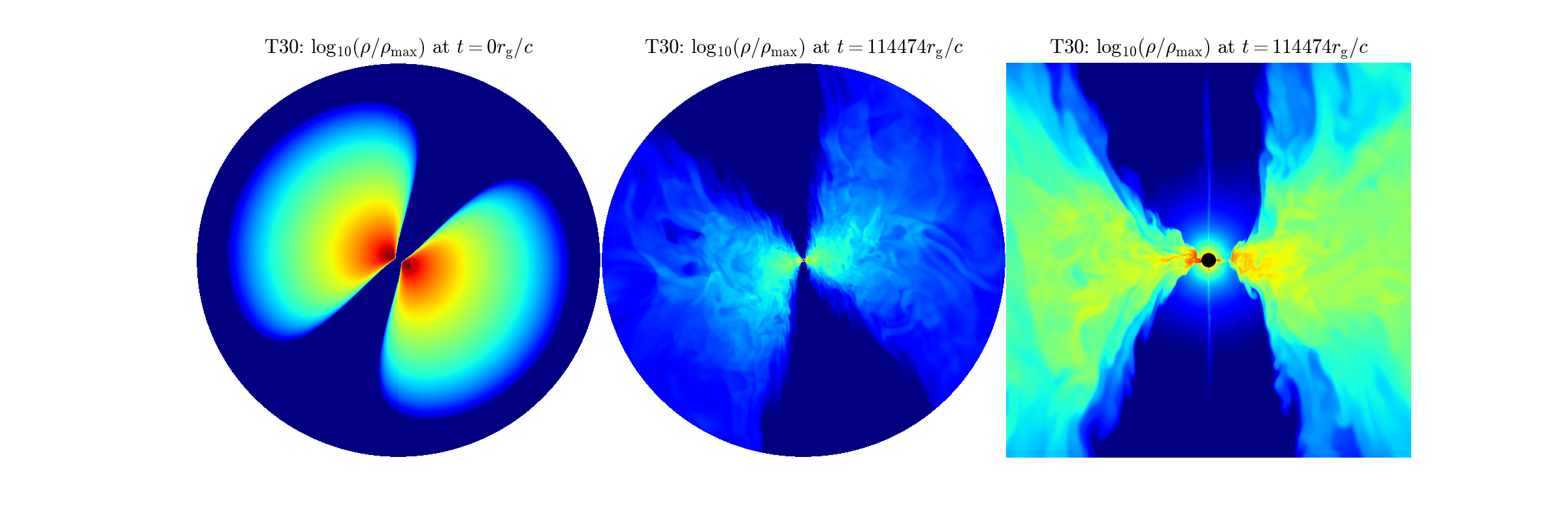}
    \caption{Initial and final states of a magneto-spin aligned accretion flow. We show vertical cross sections of the gas density ($\rho$) for a disk initially misaligned at $30^{\circ}$ with respect to the black hole spin axis at time $t=0\,\rg/c$ (left panel) and at late times (middle and right panel) The left and middle panels show the full simulation grid, out to $1000\,\rg{}$ while the right panel zooms in to the inner $80\,\rg{}\times 80\,\rg{}$ of the grid. The color bar shows five orders of magnitude in density.}
    \label{fig:T30}
\end{figure*}

Outflowing collimated streams of magnetized plasma moving at relativistic speeds, called jets, are ubiquitously observed around compact objects, like black holes (BHs) and neutron stars. Owing to the enormous energy they carry, jets deposit their mechanical energy as they travel outward from the BH and can significantly heat up interstellar medium (ISM) gas on interaction, forming a crucial component of the BH-galaxy feedback cycle \citep[][]{Fabian2012, har18feed,chatterjee2019}. Further, jets could facilitate high energy particle acceleration via shocks or magnetic reconnection, and provide a viable origin for cosmic rays \citep[][]{Caprioli:2015,sironi2015,Matthews:2018,Crumley:2019,Blandford:2019}.

Currently, the most plausible explanation for the high energy content of jets comes from the \citet[][]{bz77} (hereafter BZ) mechanism where BH inertial frame-dragging forces magnetized gas to corotate with the BH spin. This results in azimuthal twisting of vertical fields that produces an outward vertical pressure gradient and propels gas into a collimated outflow, forming jets. The BZ jet power depends on both the magnetic fields supplied to the BH as well as the BH spin $a$, and is given as $P_{\rm BZ}=(k \phi_{\rm BH}^2\Omega_{\rm H}^2) \dot{M}_{\rm BH}c^2$ \citep[][]{mck04,tch10a,Chatterjee:2023_JP}, where $\phi_{\rm BH}$, $\Omega_{\rm H}\,\,(=a/2r_{\rm H})$ and $\dot{M}_{\rm BH}$ are the dimensionless magnetic flux at the BH event horizon ($r=r_{\rm H}$), horizon angular frequency and mass accretion rate. The proportionality constant $k$ depends on the shape of the magnetic field lines threading the horizon \citep[][]{tch10a}. 

The BH spin dependence is particularly interesting as the maximum possible $\phi_{\rm BH}$ varies not only with the spin magnitude but also the direction, i.e., prograde or retrograde spin with respect to the accreting gas \citep{tch12proc,Tchekhovskoy:2015, Narayan:2022}. Spin values are notoriously difficult to measure due to the turbulent nature of the accretion flow in the case of single BHs and the uncertainty in progenitor spins for equal-mass binary BH mergers. While current spin measurements for stellar mass BHs cluster around high values (for a review, see \citep[][]{Reynolds:2021}), there is evidence that galaxy mergers may result in small spins \citep{volonteri05}. 

In regards to $\phi_{\rm BH}$, the BH would be fed by the ISM that usually has approximately equipartition magnetic fields ($\beta\sim1$; \citep[][]{Haverkorn:2013,Ferriere:2020}) due to galactic dynamos \citep[][]{Beck:1996, Brandenburg:2005}. Assuming a nominal value of $\sim\mu$Gauss, one could argue that sufficient magnetic flux is available to saturate the BH magnetosphere \citep[][]{narayan03}. Indeed, recent studies suggest that dynamically strong magnetic fields may regulate the accretion of gas onto supermassive BHs in low-luminosity active galactic nuclei (AGN), such as Sagittarius A* (Sgr A*) and M87 \citep[][]{EHT_M87_2019_PaperV,EHT_SgrA_2022_PaperV}. The end state of magnetic-saturation, known as the magnetically arrested disk (MAD), produces highly efficient jets, which carry away angular momentum large enough to cause BH spindown \citep{tch12proc,Narayan:2022}, especially for near- and super-Eddington BHs \citep{Lowell:2023, Ricarte:2023}. However, for the sub-Eddington accretion regime, conditions for the development of the MAD state and powerful jets appear optimal.

Usually, general relativistic magnetohydrodynamics (GRMHD) simulations of accreting BHs assume that the rotation axis of the accretion disk and the BH spin are either aligned or antialigned. However, AGN disks may be born misaligned, as is found in galaxy simulations that resolve AGN disk formation \citep{anglesalcazar_2017_FIRE,hopkins:FORGE2,kaaz:2024b} and is evidenced by subparsec scale maser emission that is consistent with warps \citep{greenhill_2003,zhao_2018,zaw_2020}, most notably in NGC 4258 \citep{miyoshi_1995,greenhill_1995}. Allowing a tilt between the disk and the BH spin axes introduces interesting effects in the global disk structure such as radial tilt oscillations \citep{papaloizou95}, Lense-Thirring (LT) \citep[][]{lense18} precession, Bardeen-Petterson (BP) alignment \citep[][]{bardeen75} and disk tearing \citep[][]{Nixon:2012,liska:2021_tearing}. Indeed, due to these additional effects, misaligned disks have been used to study quasiperiodic oscillations (QPOs) in lightcurves of BH X-ray binaries (BHXRBs; for a review, see \citep[][]{Ingram:2019}) and to interpret near-horizon images of Sgr A* and M87 \citep{Dexter:13,Chatterjee_2020, White_2020_tiltedimages, EHT_SgrA_2022_PaperV, Chatterjee:2024} with recent observations showing hints of periodicity in the M87 jet position angle \citep{Cui:2023} . 

Misaligned accretion onto the BH occurs via plunging streams of gas \citep{fragile07}, possibly driven by nozzle shocks (as seen in thin disks, \citep[][]{Kaaz:2023,kaaz:2024a}), in contrast to the near-axisymmetric accretion seen for aligned BH-disk systems. This effect hinders the advection of magnetic fields close to the BH, reducing the horizon magnetic flux \citep{Chatterjee_2020}. However, when supplied with MAD-level magnetic flux, the resulting jets can become powerful enough to force the disk to align, an effect known as magneto-spin alignment \citep{mckinney_2013, polko17, Ressler:2021}. The conditions that lead to magneto-spin alignment are yet to be understood clearly, especially the relation between disk alignment, BH spin and jet power. Further, it is unknown whether disk alignment is an inevitable consequence of the MAD state, which carries significant implications for galaxy evolution.

In this work, we simulate a wide variety of misaligned disks targeting the MAD state using the GRMHD code \hammer{}. In Sec.~\ref{sec:code}, we describe our numerical choices and the model setup. Section~\ref{sec:results} is broken into four parts, namely moderately and extremely misaligned disk models, the magneto-spin alignment effect, and outflow powers for tilted disks. We discuss observational implications of our results in Sec.~\ref{sec:discussion}, namely, shocks via disk-jet interactions, BH jet precession for sub-Eddington accretion flows and periodicity in jet observations. We provide a summary of our findings in Sec.~\ref{sec:summary}.

\begin{table*}
\centering
\renewcommand{\arraystretch}{1.3}
\begin{tabular}{|c | c c c c c | c c | c c c|}
\hline
\vspace*{0mm}

Tilt & Magnetic & BH & $\mathcal{T^{\circ}_{\rm disk}}$ & Resolution & $t_{\rm sim}$ & $\langle\phi_{\rm BH}\rangle$ & $\langle\eta_{\rm out}\%\rangle$ & $\langle\mathcal{T}^{\circ}_{\rm disk}\rangle$ & $\langle\mathcal{T}^{\circ}_{\rm disk}\rangle$ & $\langle\mathcal{T}^{\circ}_{\rm jet}\rangle$\\

model & field & spin & (initial) & $N_{r}\times N_{\theta}\times N_{\varphi}$ & $[10^4 \rg/c]$ &  \multicolumn{2}{c|}{($r=r_{\rm h}$)}  & ($10\rg$) & ($150\rg$) & ($10\rg$) \\

\hline
\hline
\multicolumn{11}{|c|}{Fiducial models: Tilted MADs} \\
\hline
\ta & MAD & 0.9375 & $0^{\circ}$  & $580\times288\times512$ & 8.27 & 61.2 & 190.5 & 2.7 & 1.8 & $5.61 \pm 2.56$\\
\tb & MAD & 0.9375 & $15^{\circ}$ & $580\times288\times512$ & 13.4 & 60.4 & 191.4 & 2.62 & 8.07 & $5.72 \pm 3.84$\\
\tc & MAD & 0.9375 & $30^{\circ}$ & $580\times288\times512$ & 11.4 & 59.9 & 187.8 & 2.83 & 14.9 & $7.06 \pm 4.39$\\
\td & MAD & 0.9375 & $45^{\circ}$ & $580\times288\times512$ & 12.9 & 62.2 & 198.2 & 2.61 & 23.2 & $6.46 \pm 2.95$\\
\te & MAD & 0.9375 & $60^{\circ}$ & $580\times288\times512$ & 12.4 & 60.8 & 190.5 & 2.5 & 27.4 & $5.26 \pm 2.11$\\
\tf & MAD & 0.9375 & $75^{\circ}$ & $580\times288\times512$ & 9.7 & 18 & 3.139 & 78.2 & 74.8 & $76.7 \pm 6.79$\\
\tg & MAD & 0.9375 & $90^{\circ}$ & $580\times288\times512$ & 10.1 & 19.1 & 4.257 & 80.6 & 71.5 & $75.8 \pm 8.81$\\
\ti & MAD & 0.9375 & $90^{\circ}$ & $696\times576\times576$ & 7.33 & 18 & 4.66 & 84.2 & 77.1 & $80.2 \pm 5.55$\\
\tj$^{\dagger}$ & MAD & 0.9375 & $90^{\circ}$ & $1160\times192\times384$ & 10.2 & 16.8 & 18.38 & 58.8 & 81 & $21.7 \pm 15.9$\\

\hline
\hline
\multicolumn{11}{|c|}{Tilted MAD models with alternate black hole spin} \\
\hline

\texttt{T30A5}   & MAD &     0.5 & $30^{\circ}$ & $580\times288\times512$ & 5 & 57.4 & 26.56 & 11.4 & 25.1 & $12.1 \pm 3.24$ \\
\texttt{T30A5M}  & MAD &    $-0.5$ & $30^{\circ}$ & $580\times288\times512$ & 5 & 33.1 & 6.136 & 38.5 & 31.3 & $31.5 \pm 3.89$\\
\texttt{T30A93M} & MAD & $-0.9375$ & $30^{\circ}$ & $580\times288\times512$ & 5 & 19.7 & 8.084 & 44.1 & 36.2 & $32.9 \pm 6.23$ \\

\hline
\hline
\multicolumn{11}{|c|}{Tilted Near-MAD models based on \cite{liska_tilt_2018} and \cite{Chatterjee_2020}} \\
\hline

\texttt{T0-S}      & Strong &  0.9375 &  $0^{\circ}$ & $448\times144\times240$ & 13.2 & 52.8 & 140.4 & 3.57 & 1.84 & $5.77 \pm 2.28$\\
\texttt{T30-S}     & Strong &  0.9375 & $30^{\circ}$ & $448\times144\times240$ & 11.9 & 46.3 & 102 & 20.7 & 21.8 & $18.2 \pm 2.99$\\
\texttt{T60-S}     & Strong &  0.9375 & $60^{\circ}$ & $448\times144\times240$ & 14.8 & 33.3 & 47.5 & 42.7 & 41.8 & $33.3 \pm 3.23$\\
\texttt{T30A5-S}   & Strong &     0.5 & $30^{\circ}$ & $448\times144\times240$ & 11 & 49.9 & 21.76 & 24 & 25.2 & $23.9 \pm 3.36$\\
\texttt{T30A93M-S} & Strong & $-0.9375$ & $30^{\circ}$ & $448\times144\times240$ & 5 & 12.7 & 7.309 & 22.6 & 33.6 & $23.9 \pm 1.42$\\

\hline
\hline
\multicolumn{11}{|c|}{Tilted BH models for alternate initial conditions (See Sec.~\ref{sec:accretion_history})} \\
\hline

\tiltBHa & MAD & 0.9375 & $0^{\circ}$ & $580\times192\times384$ & 5 & 54.3 & 143.3 & 3.02 & 6.37 & $7.05 \pm 3.92$\\
\tiltBHb & MAD & 0.9375 & $\cdots$ & $580\times192\times384$ & 5 & 25 & 30.1 & 42.6 & 66.8 & $46.8 \pm 20.9$\\
\tiltBHc & MAD & 0.9375 & $\cdots$ & $580\times192\times384$ & 5 & 23.2 & 15.21 & 28 & 27.5 & $35.5 \pm 12$\\
\tiltBHd & MAD & 0.9375 & $\cdots$ & $580\times192\times384$ & 5 & 9.25 & 0.8024 & 26.4 & 26.9 & $34.5 \pm 19$\\
\hline

\end{tabular}
\caption{Summary table for all numerical simulations used in this work. We mention the model names, the magnetic field configuration for the initial disk (either a MAD or a strongly magnetized disk), initial disk tilt angle ($\mathcal{T}_{\rm disk}$), simulation grid resolution and simulation end time $t_{\rm sim}$. We include time-averaged simulation quantities: mass accretion rate $\dot{M}$, dimensionless magnetic flux $\phi_{\rm BH}$, outflow efficiency $\eta_{\rm out}$, and the final disk tilt angles $\mathcal{T}_{\rm disk}$ at $10\,\rg$ and $150\,\rg$ (indicating the inner and outer disks). We average the simulation quantities over the final $10^4\,r_{\rm g}/c$ for each simulation. $^{\dagger}$ Note that \texttt{T90L} has an outer boundary of $10^6\,\rg{}$. 
} 
\label{tab:tiltmod}
\end{table*}

\section{Numerical method and Model Setup}
\label{sec:code}

We use the GPU-accelerated code \textsc{h-amr} \citep{liska_hamr:2022} that solves the GRMHD equations in a fixed Kerr spacetime. We use a uniform grid in logarithmic spherical polar Kerr-Schild coordinates ($t$, $\log r$, $\theta$, $\varphi$). We adopt geometrical units, i.e. $G=c=1$ in which the gravitational radius, $\rg=GM_{\rm BH}/c^2$, equals the BH mass, $M_{\rm BH}$. Our simulations have an effective grid resolution of $N_r\times N_{\theta}\times N_{\varphi}=580\times 288\times512$. The grid extends from $r\in [1.1525,1000]\rg$, $\theta\in [0,\pi]$ and $\varphi\in [0,2\pi]$. In order to speed up our simulations, we derefine the $\varphi$ resolution by a factor of 4 near the polar axis. We checked the evolution of non-spinning BHs for an aligned disk (i.e., the disk midplane coincides with $\theta=\pi/2$) and a highly misaligned disk, where the disk midplane coincides with the polar axis (see Appendix~\ref{sec:zero_spin}) and found that the disk evolution is not affected when crossing the polar axis. We adopt outflowing radial boundary conditions (BCs), transmissive polar BCs and periodic BCs in the $\varphi$ direction \citep{liska_tilt_2018,liska_hamr:2022}. 

We simulated a suite of accretion systems with the disk midplane misaligned at a variety of tilt angles, $\mathcal{T}_{\rm disk}=(0,\pi/2)$, set by the BH and disk midplanes). Table~\ref{tab:tiltmod} shows a summary of our model set, in which we follow the following naming convention: \texttt{Tx}$<$suffix$>$ where ``\texttt{x}'' denotes the initial disk tilt angle, and the suffix gives additional information about the model. We initialize the disk as an equilibrium hydrodynamic torus \citep{Fishbone:76} around a BH with a fiducial spin value of $a=0.9375$. We set up a standard MAD magnetic field configuration \citep[][]{Chatterjee:2022}. For our gas thermodynamics, we assume an ideal gas equation of state with an adiabatic index of $13/9$. The magnetic field strength is normalized by setting the ratio of maxima between the gas and magnetic pressures to 100. For tackling the evacuated region in the jet funnel, we adopt the density floor injection scheme of \citet{Ressler:17} when the magnetization exceeds 20. Figure~\ref{fig:T30} shows the initial disk setup, and its late-stage evolved state.

\section{Results}
\label{sec:results}

\begin{figure*}
    \centering
    \includegraphics[width=\textwidth,trim= 0 0 0 0, clip]{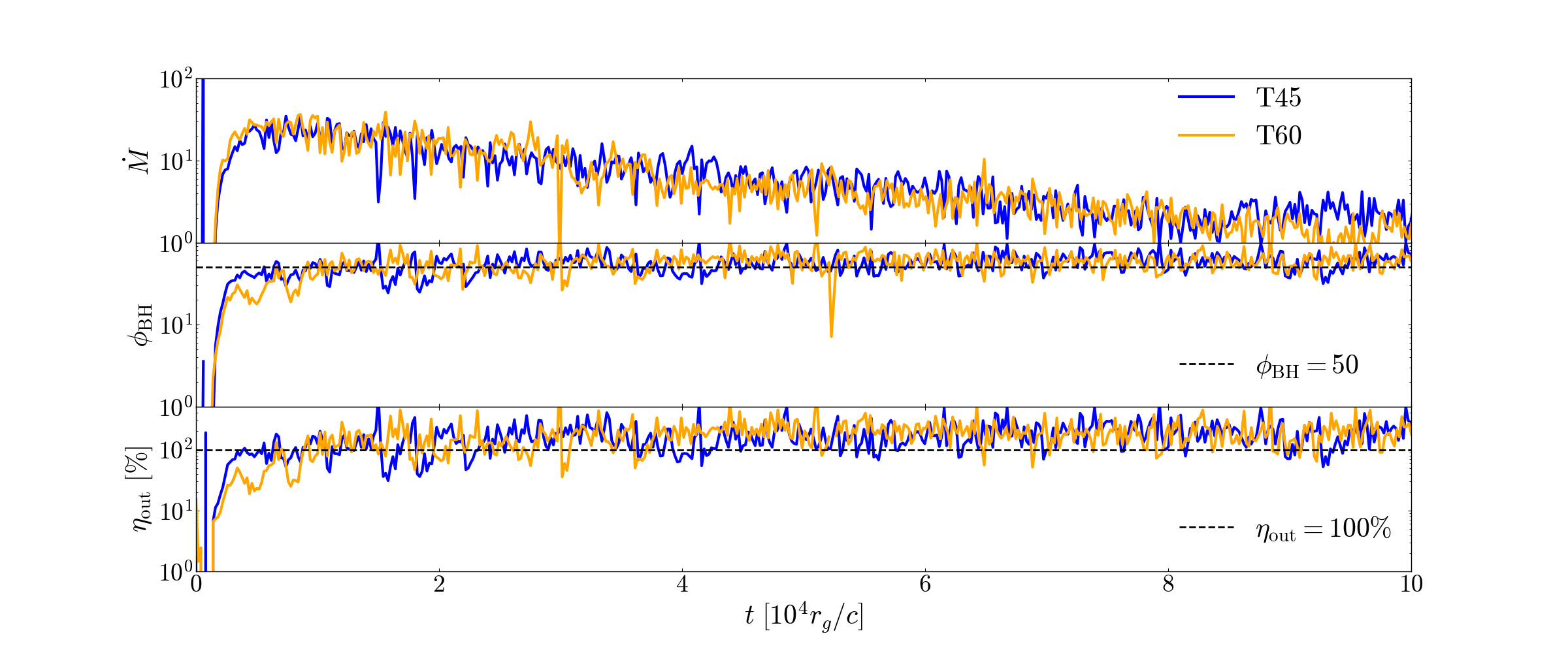}
    \includegraphics[width=\textwidth,trim= 0 0 0 0, clip]{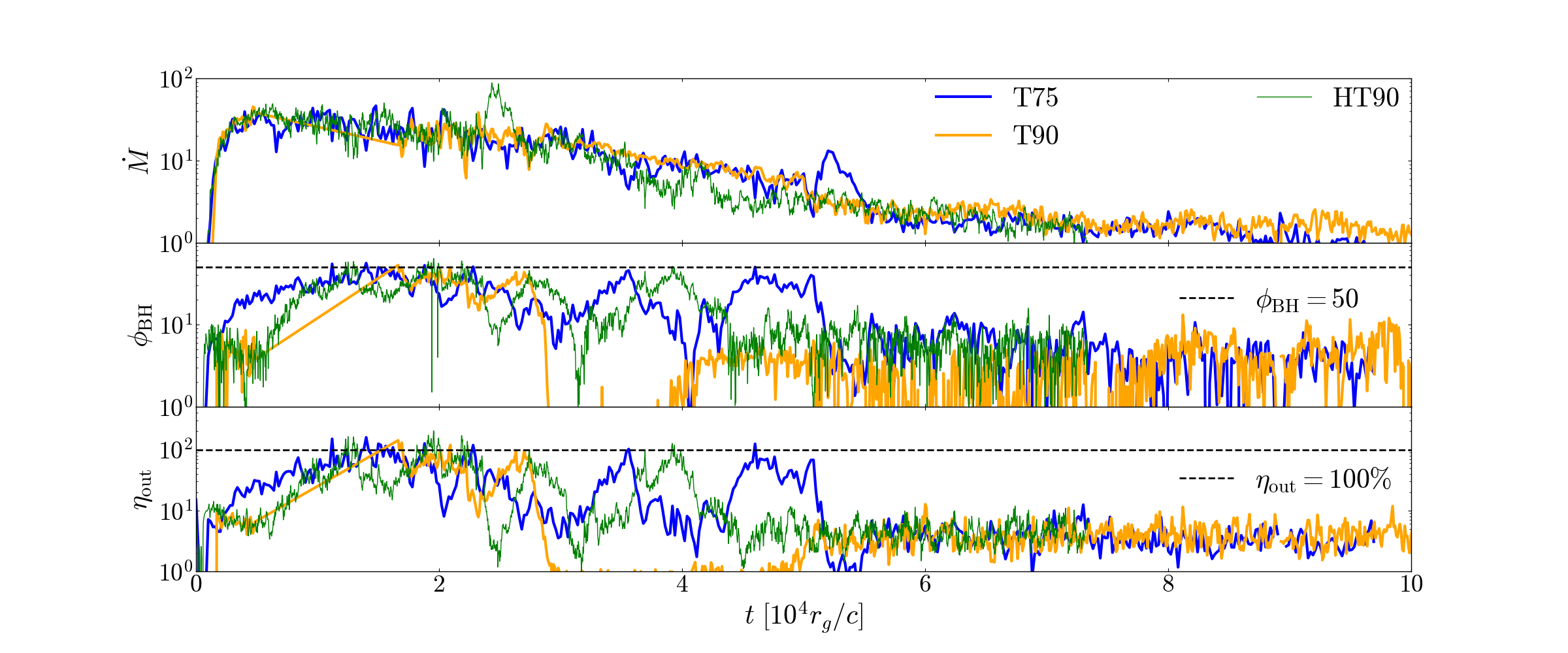}
    \caption{Disks with initial tilt angles smaller than $60^{\circ}$ develop the MAD state, while highly tilted disks undergo rapid transitions in magnetic flux and outflow power. We show the temporal evolution of the accretion rate $\dot{M}$ (in code units), the dimensionless magnetic flux $\phi_{\rm BH}$ and the outflow efficiency $\eta_{\rm out}=P_{\rm out}/\dot{M}$ for moderately-misaligned disks \td{} and \te{} (top) and highly-misaligned disks \tf{} and \tg{} (bottom). We also include \ti{}, a higher resolution version of \tg{}.  
    }
    \label{fig:time}
\end{figure*}

\begin{figure}
    \centering
    \includegraphics[width=\columnwidth,trim= 0 0 0 0, clip]{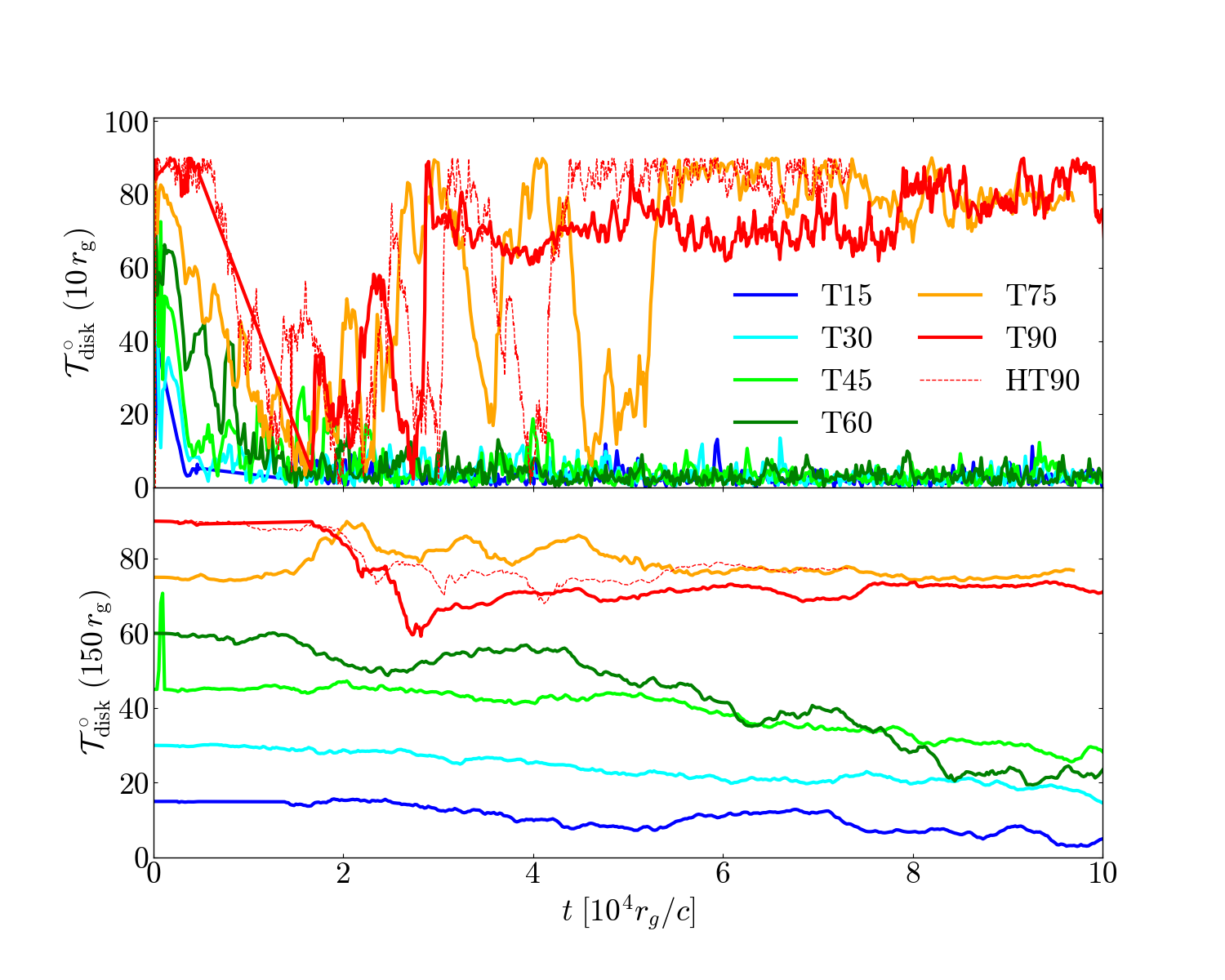}
    \caption{We show the disk tilt angle for our fiducial tilted MAD simulations at $10\,\rg$ (top) and $150\,\rg$ (bottom). Disks that develop the MAD state (Fig.~\ref{fig:time}) exhibit strong alignment with the BH spin while highly tilted disks show periods of alignment for the inner disk when the magnetic flux is large enough to produce a jet. 
    }
    \label{fig:tilt}
\end{figure}

\begin{figure}
    \centering
    \includegraphics[width=\columnwidth,trim= 0 0 0 0, clip]{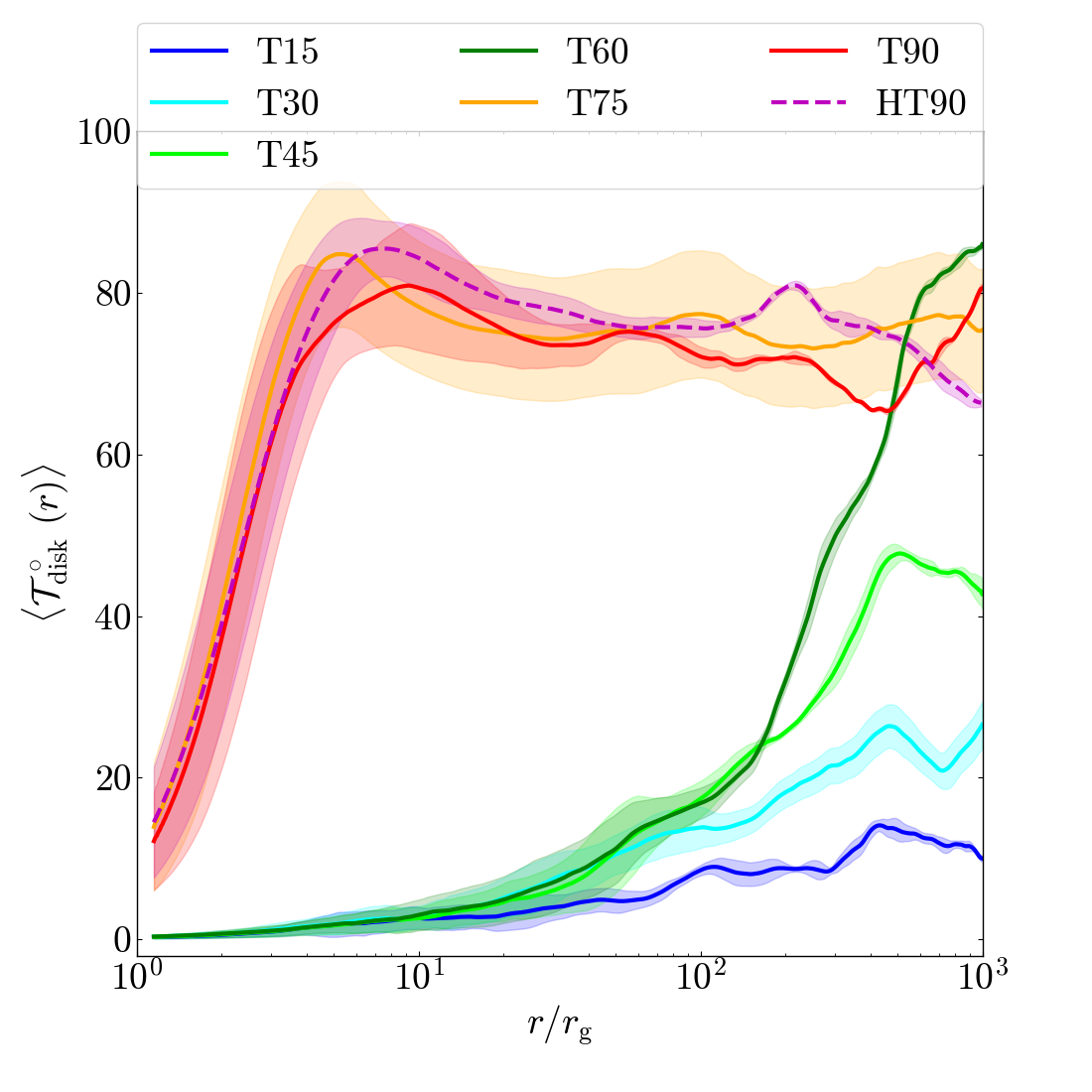}
    \caption{Radial profiles of the disk tilt angle for our fiducial models, time-averaged over the final $20,000\,\rg/c$. Moderately tilted disks show strong alignment at least to radius $r\gtrsim10\,\rg$ while highly tilted disks exhibit radial tilt oscillations as expected for weakly magnetized tilted disks \citep[][]{liska_tilt_2018}.
    }
    \label{fig:tilt_rad}
\end{figure}

Our first set of fiducial model set comprises seven GRMHD simulations of varying initial disk inclinations evolved to $\gtrsim 0.8-1 \times 10^5~\rg/c$. At these late times, the radial profiles of different disk properties, such as the disk tilt angle $\mathcal{T}$, become roughly independent of time, and so the bulk of the disk achieves quasi$-$steady state behavior. Table~\ref{tab:tiltmod} shows a summary of the simulation properties and some relevant time-averaged quantities for our models. 

\subsection{Moderately misaligned flows}
\label{sec:moderate}

First, we discuss the moderately misaligned disk simulations (models \texttt{T15} to \texttt{T60}). The top panel of Fig.~\ref{fig:time} shows the mass accretion rate $\mdot=-\iint\rho u^r\sqrt{-g} d\theta d\varphi$ of the incoming flow, the dimensionless magnetic flux $\phi_{\rm BH}=0.5\sqrt{4\pi/\mdot}\iint_{r=r_{\rm h}} |B^r|\sqrt{-g} d\theta d\varphi$, evaluated at the event horizon $r_{\rm h}=\rg(1+\sqrt{1-a^2})$, and the outflow power efficiency $\eta_{\rm out}=P_{\rm out}/\mdot_{r_0} c^2=(\mdot-\dot{E})/\mdot_{r_0}$ (taking $c=1$), for two of the misaligned models, \td{} and \te{}. Here we use the energy flux $\dot{E}=\iint T^r_t\sqrt{-g} d\theta d\varphi$, where $T^{\mu}_{\nu}$ is the stress-energy tensor. Here $g$ is the metric determinant. We calculate both $\mdot$ and $P_{\rm out}$ at $r=5\,\rg$ so as to avoid contamination due to density floors. The power efficiency is the outflow power normalized by the instantaneous $\mdot$ at $r_0=5\,\rg$.

In Fig.~\ref{fig:time}, despite the difference in the initial disk inclination ($45^{\circ}$ versus $60^{\circ}$), the $\mdot$, $\eta_{\rm out}$ and $\phi_{\rm BH}$ of the two models evolve very similarly over time. The outflows from both models exceed efficiencies of 100\% and the accretion flow at small radii reaches magnetic flux saturation, indicated by the time-averaged magnetic flux $\langle\phi_{\rm BH}\rangle\gtrsim 50$ and the clear presence of dips in the magnetic flux and accretion rate due to magnetic flux eruptions in the time plots. Indeed, models \texttt{T15} to \texttt{T60} all develop the MAD state ($\phi_{\rm BH}\gtrsim 50$) and exhibit highly efficient outflows, similar to the nontilted \texttt{T0} MAD model (see Table~\ref{tab:tiltmod}), and typical of geometrically thick MAD flows around rapidly-spinning prograde BHs (both $\phi_{\rm BH}$ and $\eta_{\rm out}$ depend on the BH spin and the disk scale-height; \citep{tch12proc, Narayan:2022, Lowell:2023}).   

For all of the moderately tilted disk models, as the accretion flow develops the MAD state, the strongly magnetized jet generates enough electromagnetic torque to force the inner disk flow to align with the black hole spin axis (see Appendix~\ref{sec:accretion_history}. The ``magneto-spin alignment'' of the disk results in near-zero values of the disk tilt angle close to the BH (shown at a radius of $10~\rg$ in Fig.~\ref{fig:tilt}, top panel) over time. On the other hand, the bottom panel of Fig.~\ref{fig:tilt} shows that at larger distances from the BH (specifically $r=150\,\rg$ in the plot), the outer disk undergoes relatively smaller levels of alignment. Figure~\ref{fig:tilt_rad} shows the radial profile of the disk tilt angle, averaged over the final $20000\,\rg/c$. The moderately misaligned disks exhibit near-perfect alignment with the BH spin axis at least out to $10\,\rg$, beyond which the flow remains misaligned. It is possible that the alignment radius increases out to large radii over several viscous timescales, depending on whether the flow angular velocity remains sub-Keplerian \citep[][]{mckinney_2013}. We also note that the jet torque roughly realigns the inner disk to its final orientation within the first $10^4\,\rg/c$ of its evolution, much faster than the BP alignment timescales.

\begin{figure*}
    \centering
    \includegraphics[width=0.49\textwidth,trim= 2cm 2cm 2.5cm 1cm, clip]{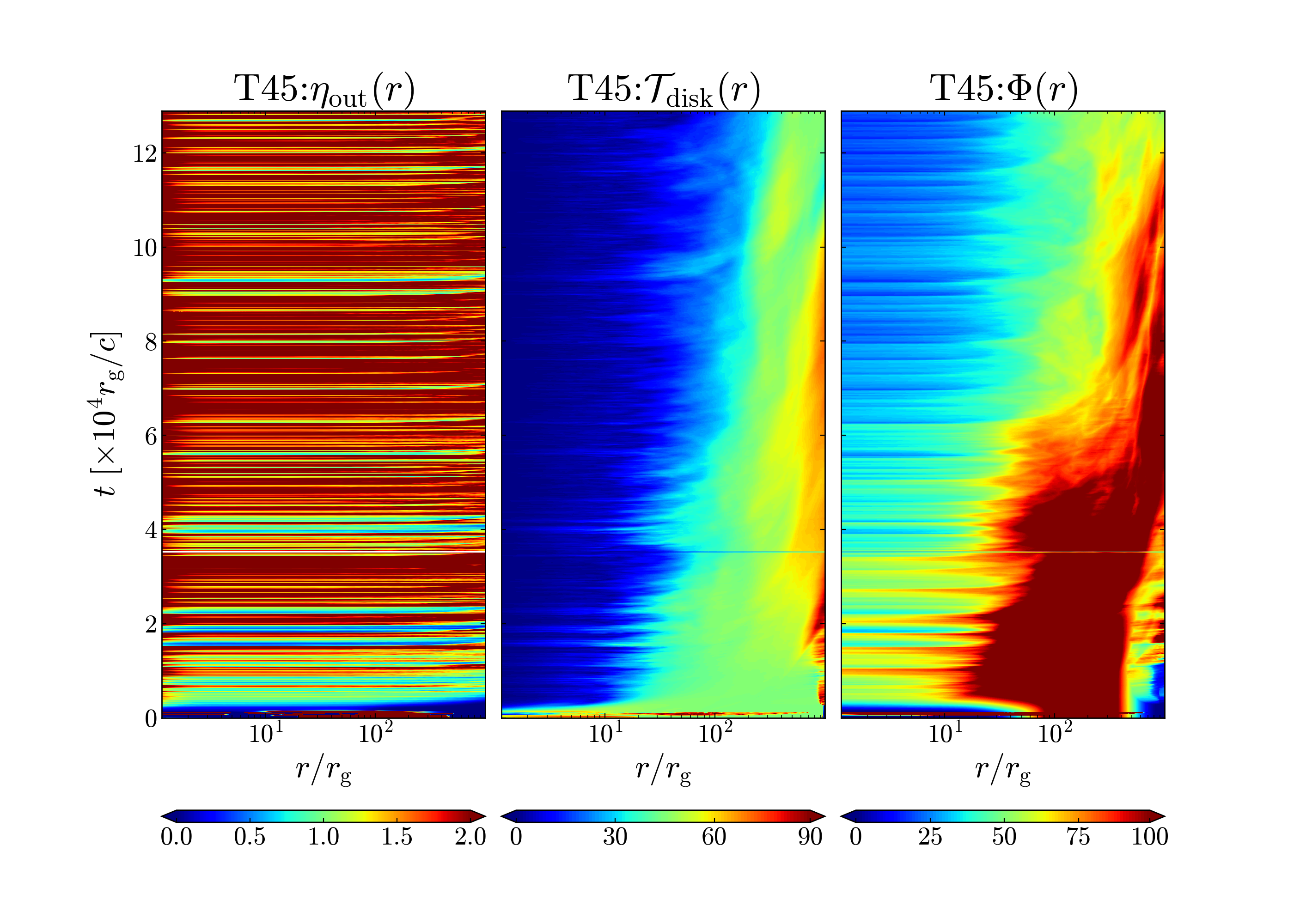}
    \includegraphics[width=0.49\textwidth,trim= 2cm 2cm 2.5cm 1cm, clip]{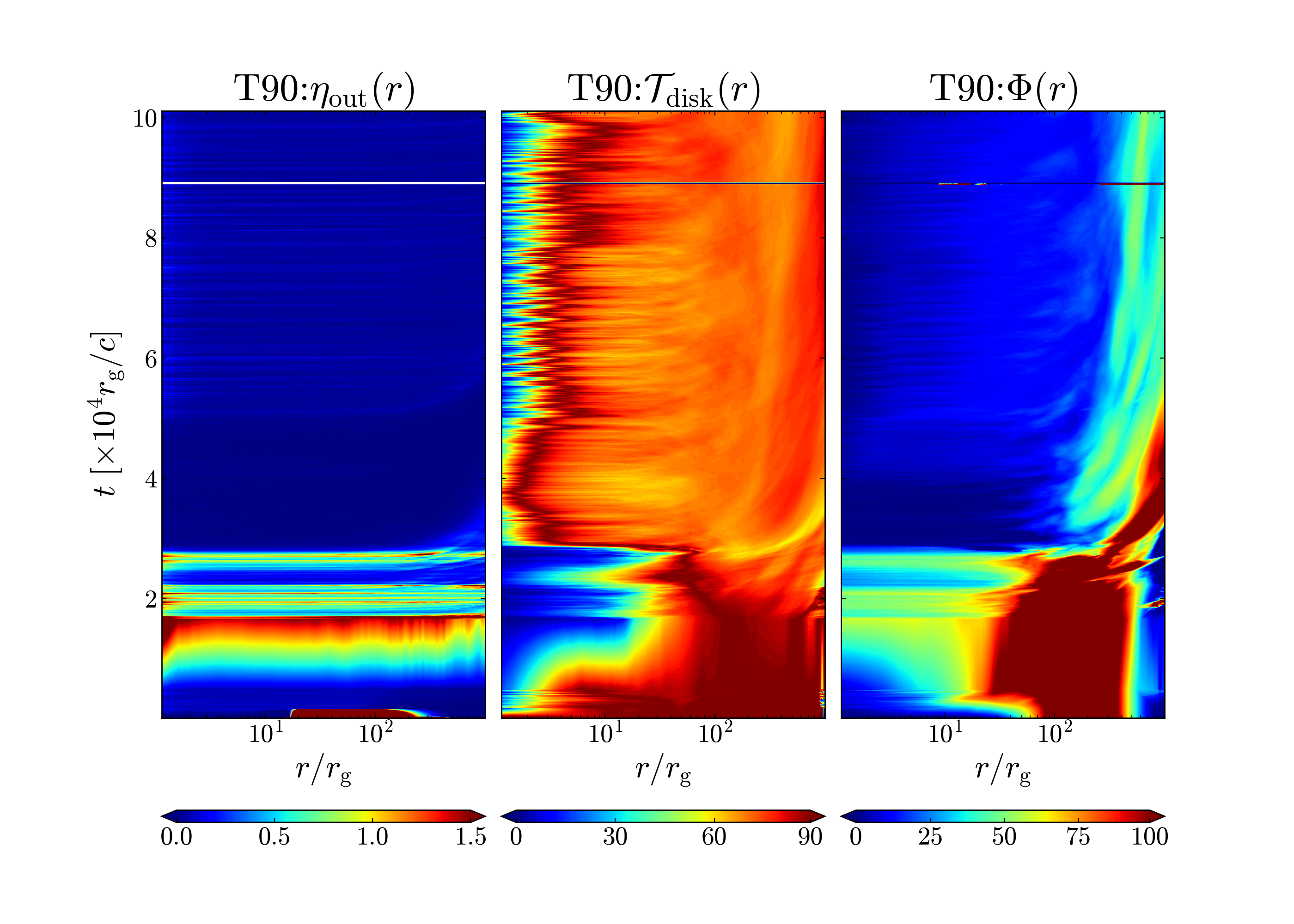}
    \includegraphics[width=0.49\textwidth,trim= 2cm 2cm 2.5cm 1cm, clip]{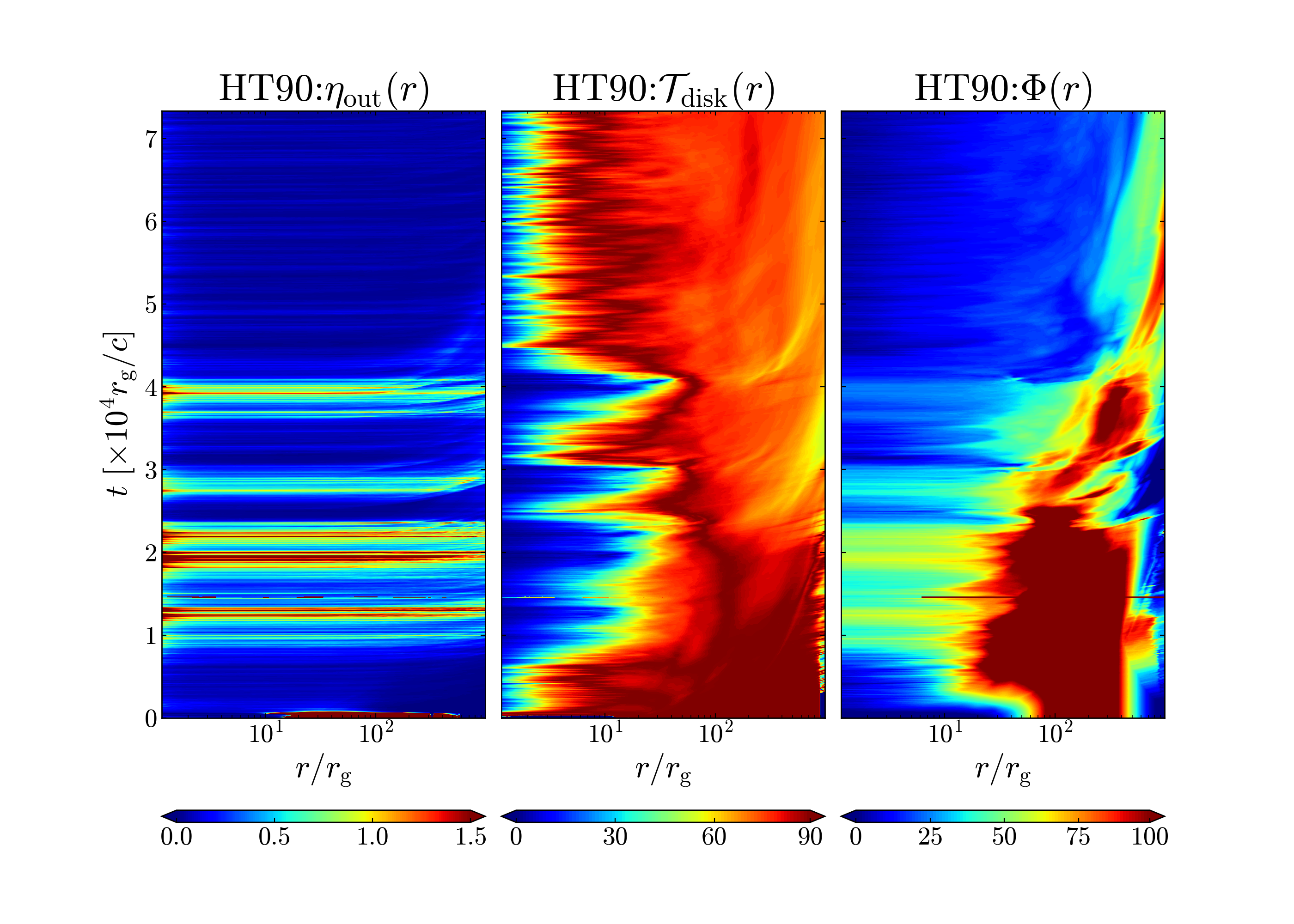}
    \includegraphics[width=0.49\textwidth,trim= 2cm 2cm 2.5cm 1cm, clip]{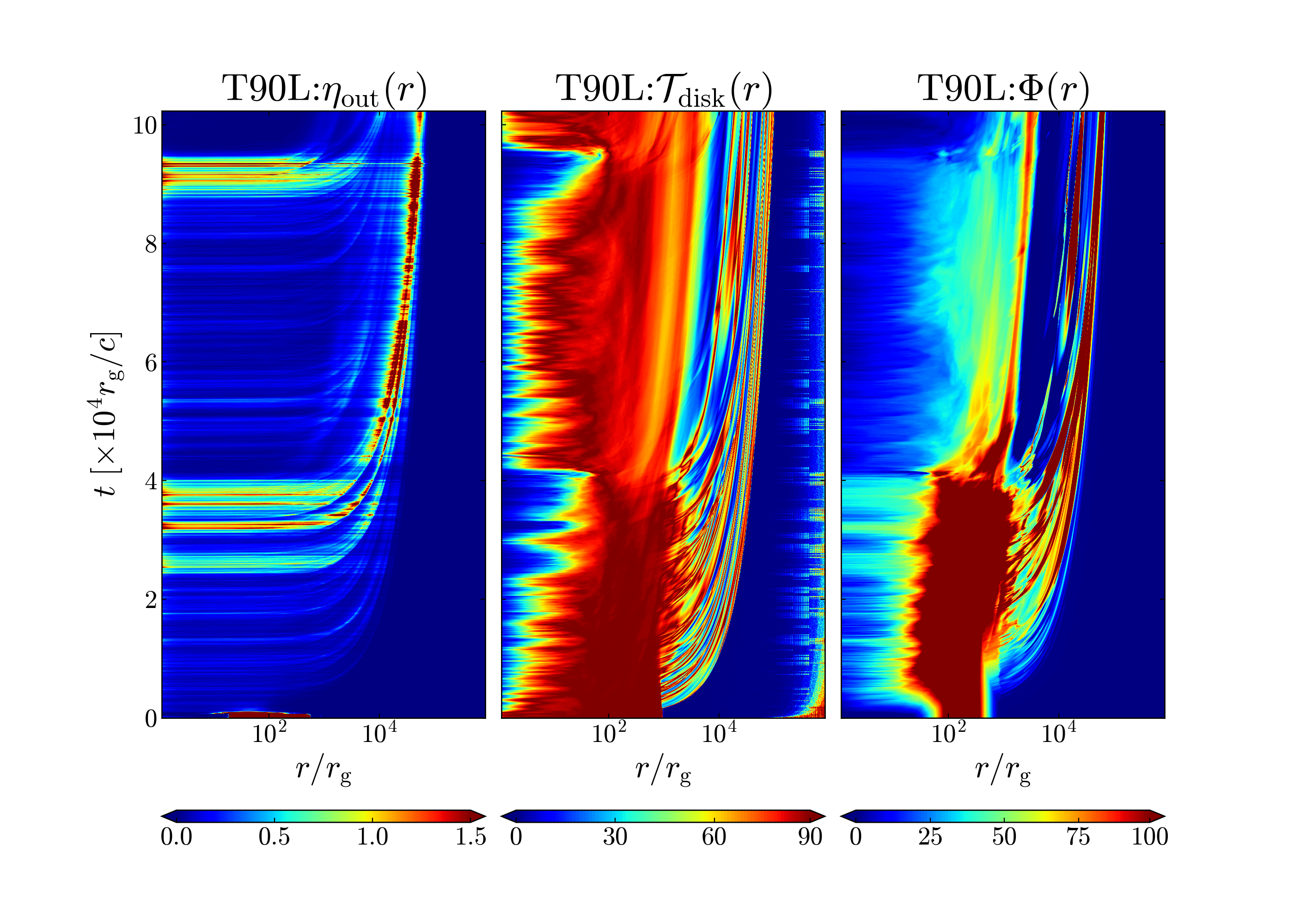}
    \caption{We show the time and radial evolution of the outflow power $\eta_{\rm out}$, and its relation to the disk tilt angle $\mathcal{T}_{\rm disk}$ and the enclosed magnetic flux $\Phi$ in the disk for models \texttt{T45}, \texttt{T90}, \texttt{T90H} and \texttt{T90L}. We see that intermittent jets, indicated by periods of large outflow power, force the disk to align with the BH's spin axis. However, due to strong collisions between the disk and the jets, magnetic flux gets ejected out to large radii, consequently resulting in weakened outflows with $\eta_{\rm out}\ll 10\%$. On the other hand, for the moderate tilt case, $\texttt{T45}$, we have a steady and powerful jet with $\eta_{\rm out}\gtrsim 150-200\%$ where the efficiency is approximately constant in time and distance from the BH.}
    \label{fig:phibh}
\end{figure*}

\subsection{Extremely misaligned disks and intermittent jets}
\label{sec:extreme}
 
The extremely tilted disk models \texttt{T75} and \texttt{T90}, with initial disk misalignment angles of $75^{\circ}$ and $90^{\circ}$, respectively, evolve entirely differently from the moderate spin cases. Focusing on the late time-averaged properties first, these two models do not end up in a quasisteady MAD state, but instead exhibit a weakly magnetized accretion flow. The models show a dimensionless magnetic flux $\langle\phi_{\rm BH}\rangle \lesssim 6$ (Fig.~\ref{fig:time} and Table~\ref{tab:tiltmod}) along with weakly powered jets with efficiencies of a few percent. In these models, magneto-spin alignment of the disk does not seem to occur: the tilt angle of the inner disk stabilizes to $\sim80^{\circ}$ (from Fig.~\ref{fig:tilt}). Indeed, the radial profile of the disk tilt angle (Fig.~\ref{fig:tilt_rad}) exhibits radial tilt oscillations reminiscent of relatively weakly magnetized misaligned accretion flows (as seen in, e.g., \citep{fragile07, liska_tilt_2018, White_2019, Chatterjee_2020}). The strong disk alignment only occurs within a few gravitational radii from the BH. 

Despite starting with physical conditions conducive to forming a MAD state, why do our extremely misaligned disks somehow appear similar to their weakly magnetized cousins? The answer hides in their time evolution. Figure~\ref{fig:time} shows that both \texttt{T75} and \texttt{T90} reaches $\phi_{\rm BH}\approx 50$ along with relatively high outflow efficiencies (indicative of powerful jets) around $2\times10^4\,\rg/c$. Just beyond this point, at $t\sim2 - 6\times10^4\,\rg/c$, both $\phi_{\rm BH}$ and $\eta_{\rm out}$ show rapid variability and eventually settle to low values that are representative of weakly magnetized disks. The outflow power profile follows the magnetic flux as the jet is powered via the \citet{bz77} mechanism: $\eta_{\rm jet}\propto a^2\phi_{\rm BH}^2$. The temporal profile of the inner and outer disk tilt angles in Fig.~\ref{fig:tilt} replicates the highly variable behavior of $\eta_{\rm out}$ since whenever the jet power increases, the torque on the disk due to the jet becomes large enough to activate magneto-spin alignment, decreasing the disk tilt angle.  

To make sure that our results are independent of grid resolution, we ran a higher resolution version of \texttt{T90}, called \texttt{T90H} with double the number of grid cells in the $\theta$ direction to ensure that the accretion flow is very well resolved even near the grid polar axis region. We find that the fiducial and high resolution models evolve similarly in time and show converged disk tilt angle profiles (Figs.~\ref{fig:time}, \ref{fig:tilt}, and \ref{fig:tilt_rad}). However, we note that transient jetted phase of \texttt{T90H} shows more high-power ejection events as is evident from the presence of peaks in $\eta_{\rm out}$ at $t\sim3-5\times10^4\,\rg{}/c$ in Fig.~\ref{fig:time}. Additionally, we have performed grid tests for Schwarzschild BHs with disks aligned with the $\theta=0$ and $\theta=\pi/2$ planes and found that our disks evolve similarly, providing confidence in our base grid resolution (see Appendix~\ref{sec:zero_spin}).

What causes the variability in the magnetic flux and further, why does it settle down to such low values at late times? To answer this question, in Fig.~\ref{fig:phibh}, we construct space-time evolution plots of the outflow power efficiency $\eta_{\rm out}$, the disk tilt angle $\mathcal{T}_{\rm disk}$, and the disk poloidal magnetic flux, defined as $\Phi(r)={\rm max}_{\Tilde{\theta}} \iint_{0}^{\Tilde{\theta}} B^r(r,\Tilde{\theta},\Tilde{\varphi})\sqrt{-g} d\Tilde{\theta} d\Tilde{\varphi}$ \citep[][]{Liska:19}. Here, the ``tilde'' denotes rotated spherical polar coordinates such that the coordinate frame is aligned with the disk angular momentum vector at every radius $r$. 

The space-time plots show that jet production in these models is intermittent and the disk orientation undergoes rapid changes depending on the strength of the magnetic flux near the BH horizon. Whenever there is enough magnetic flux on the BH horizon to launch a jet, these jets not only force the disk into alignment with the BH spin (indicated by the sharp reduction in $\mathcal{T}_{\rm disk}$ when $\eta_{\rm out}$ is large) but also inadvertently collide with the disk bulk. These jet-disk collisions push gas and magnetic flux outward, which depletes $\phi_{\rm BH}$ and reduces jet power. As the jet power reduces in strength, the angular momentum vector of the accreting gas in the inner disk switches back to that of the large scale disk. Over time, magnetic flux reaccumulates near the BH, relaunching the jets. This oscillatory behavior in the inner disk tilt angle continues until enough magnetic flux has escaped the simulation domain and the disk reaches a quasi$-$steady state warped structure reminiscent of weakly magnetized misaligned accretion flows seen in \citet[][]{Chatterjee_2020}, typically dubbed as ``Standard and Normal Evolution'' accretion flows in GRMHD literature \citep[][]{narayanSANE2012,Porth:19}. 

In both \texttt{T75} and \texttt{T90} models, any outflows, when generated, push the bulk of the available gas and magnetic flux out to large radii and eventually out of the simulation domain that extends out to $r=10^3\rg$. This behavior is confirmed by the higher resolution simulation \ti{}, increasing confidence in our results. Such a result is a feature of both the (relatively) small outer radial boundary of $r=10^3\,r_{\rm g}$ and our use of outflow radial boundary conditions that do not resupply the simulation domain with any additional gas or magnetic flux over time as opposed to Bondi-Hoyle-Lyttleton \citep[]{Hoyle:1939, bondi:1944, Edgar:2004} type models (e.g., quasi-spherical rotating inflows or wind-fed inflows; \citep{Lalakos:2022, Ressler:2023}) where a common choice is to fix the magnetic field at the outer radial boundary of the simulation domain. 

To eliminate the effect of a finite outer radial boundary, we move it to $10^6\,r_{\rm g}$ and to keep the same radial grid resolution as the \texttt{T90} model, we double the number of cells in the $r$ direction. The resulting simulation, \texttt{T90L}, exhibits the same expulsion of the magnetic flux from the inner disk due to jet-disk collisions as seen for models \texttt{T90} and \texttt{T90H} in Fig.~\ref{fig:phibh} at $t\sim 4\times10^4\, r_{\rm g}/c$. Because of the larger domain size in model \texttt{T90L}, the expelled flux remains within the simulation domain and advects inward along with the gas on a  viscous timescale and finally accumulates near the BH. This results in a restarted jet at $t\sim 9\times 10^4\, r_{\rm g}/c$. After another bout of powerful jet production, the magnetic is re-expelled and the whole cycle begins anew. This is the first demonstration of self-quenching of the MAD state and the production of restarted jets in thick accretion flows. 

\begin{figure}
    \centering
    \includegraphics[width=\columnwidth,trim= 0 0 0 0, clip]{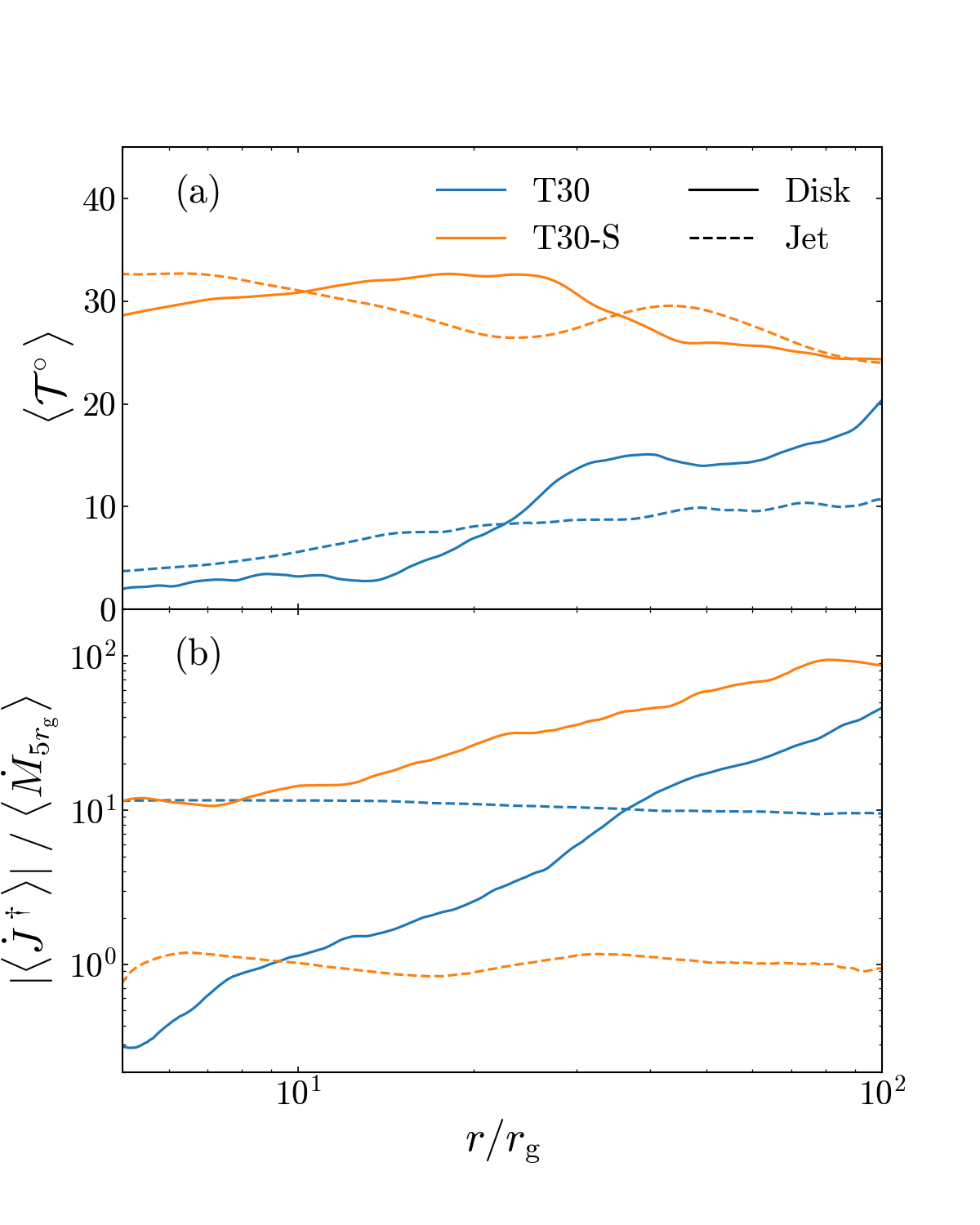}
    \caption{Jets for rapidly spinning BHs with MAD flows carry much larger angular momentum flux and deflect the disk significantly, showing the mechanism behind magneto-spin alignment. We show the (a) disk and jet tilt angle $\mathcal{T}^{\circ}$, and (b) the magnitude of the normalized angular momentum flux $\dot{J}^{\dagger}$ for two models with the same BH spin $a_*=0.9375$ and initial disk tilt angle $30^{\circ}$, a MAD (\texttt{T30}), where the jet aligns the disk, and a strong field disk (\texttt{T30-S}), where the disk orientation sets the jet direction. For MAD-level magnetic flux supply, the jet angular momentum flux (dashed lines, panel b) in the inner several $r_{\rm g}$ is much larger than the disk flux (solid, panel b), the disk tends to align. The same is also true for rapidly spinning BHs as compared to low spin BHs. To get smooth radial profiles, we performed time averaging over $1000\rg/c$ around $5\times10^4\rg/c$. The angular momentum flux is normalized by the accretion rate at $5\,\rg$.
    }
    \label{fig:angmomflux}
\end{figure}

\begin{figure}
    \centering
    \includegraphics[width=\columnwidth,trim= 0 0 0 0, clip]{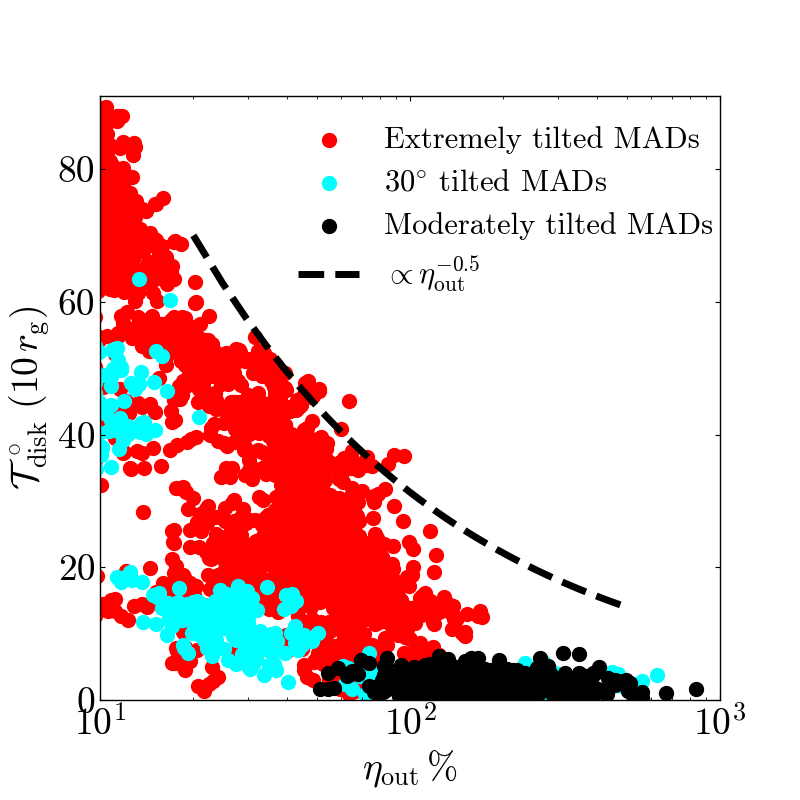}
    \caption{Magneto-spin alignment in misaligned disks: the higher the outflow power, the stronger the disk/jet alignment. We show the disk tilt angle as a function of the outflow power $\eta_{\rm out}$ during different times of evolution for our fiducial models (``Moderate'' and ``Extreme'' cases) as well as $30^{\circ}$-tilted MAD models with alternate BH spins. The magneto-spin alignment mechanism provides an upper limit on the maximum disk tilt angle for a specific jet power. For the normalization of the outflow power, we use the instantaneous accretion rate at $5\,\rg$. During periods of magnetic flux eruptions, $\mdot\rightarrow0$, resulting in instances of extreme efficiency, $\eta_{\rm out}\gtrsim 500\%$.}
    \label{fig:tdisk_vs_eta}
\end{figure}

\subsection{The magneto-spin alignment mechanism}
\label{sec:alignment}

Figure~\ref{fig:phibh} also shows the evolution of model \texttt{T45} where the disk misalignment angle is small enough for the jets to avoid head-on collisions with the disk, and magnetic flux accumulates unimpeded, leading to the MAD state. For such small misaligned disks, in previous studies of magneto-spin alignment, the alignment radius ($r_{\rm msa}$) increases with time out to a distance where the jet torque equals the disk torque \citep[][]{mckinney_2013, polko17},
\begin{equation}
    r_{\rm msa} \propto \frac{a \Phi^2}{\dot{M} v_{\varphi}}.
\end{equation}
Here, instead of examining the jet-disk torques, we focus on the angular momentum flux from the disk and the jet. We take two models with the same BH spin, but different disk magnetic flux: MAD-level (\texttt{T30}) and strong field (or near-MAD; \texttt{T30-S}). Figure~\ref{fig:angmomflux}(a) shows that the disk and the jet remain misaligned with the BH spin for \texttt{T30-S}, while for the MAD model \texttt{T30}, the disk tends to align until $r\approx 20-30\rg$ to about half of its initial disk tilt of $30^{\circ}$, while the jet remains appreciably aligned to large radii ($\mathcal{T}< 15^{\circ}$).  

To identify the alignment mechanism, we calculate the disk and jet angular momentum flux $\dot{J}^{\dagger}=\iint T^r_{\tilde{\varphi}}\sqrt{-g} d\tilde{\theta} d\tilde{\varphi}$ separately, where we rotate the coordinate-axis ($r,\tilde{\theta},\tilde{\varphi}$) as a function of radius to be aligned with their respective orientations. We do not assume the disk and jet are aligned in this calculation. The disk exhibits a net inward angular momentum flux and the jet shows outward flux for both models. Figure~\ref{fig:angmomflux}(b) shows that the disk angular momentum flux for both models obey power-laws, while the jet angular momentum flux remains roughly constant. We find that disk angular momentum flux dominates over that of the jet for the strong-field model at all radii. The jet angular momentum flux is much larger for the MAD model, resulting in a net outward angular momentum flux until $r\approx40\rg$. This shows that the jet exchanges angular momentum with the disk and deflects the disk such that the direction of the jet angular momentum changes but not its magnitude. 

We estimate the alignment radius by equating the disk and jet angular momentum flux: $\dot{J}_{\rm disk}\approx \dot{M}_{\rm BH}(r/r_0)^{\alpha}(r/\rg)^{1/2}c$ and $\dot{J}_{\rm jet}\approx \eta_{\rm out}\mdot_{\rm BH}c^2/\Omega_{\rm F}$. The jet field line angular frequency is related to the horizon angular frequency via the BZ mechanism by $\Omega_{\rm F}\approx \Omega_{\rm H}/2$ \citep[][]{kom01}, giving us:
\begin{equation}
    r_{\rm msa}/\rg \approx  \left(\frac{2\eta_{\rm out}r_0^{\alpha}}{\Omega_{\rm H}}\right)^{\frac{1}{\alpha+\frac{1}{2}}}
    \approx \left(\frac{\phi_{\rm BH}^2\Omega_{\rm H}r_0^{\alpha}}{120}\right)^{\frac{1}{\alpha+\frac{1}{2}}}.
\end{equation}
\noindent For a high-spinning BH MAD model, $\phi_{\rm BH}\simeq50$, $\Omega_{\rm H}\sim 1/2$ and $\alpha\sim1/2$. Taking a nominal value for $r_0\sim 15\rg$ (where the accretion rate roughly matches the mass outflow rate), we get an estimate for the alignment radius:
\begin{equation}
    r_{\rm msa} \approx 40\rg \left(\frac{\phi_{\rm BH}}{50}\right)^{\frac{4}{2\alpha+1}}\left(\frac{\Omega_{\rm H}}{0.5}\right)^{\frac{2}{2\alpha+1}}\left(\frac{r_0}{15r_{\rm g}}\right)^{\frac{2\alpha}{2\alpha+1}}.
\end{equation}
For weakly magnetized flows, $\phi_{\rm BH}\lesssim10$, which gives us $r_{\rm msa}\lesssim \rg$, matching our expectations for minimal alignment when magnetic fields are dynamically sub-dominant to the gas. For slowly spinning BHs, $\Omega_{\rm H}\rightarrow0$ suggesting that the disk does not align no matter how magnetized the accreting gas is. For retrograde BHs, the accretion flow is torqued to larger misalignment angles by the LT mechanism, even if the inflow has MAD-level magnetic flux as the saturated $\phi_{\rm BH}$ is much smaller compared to prograde BHs.

Our results suggest that for MAD flows around rapidly-spinning BHs, the jet significantly aligns the disk only out to the distance where the angular momentum flux magnitude carried by the jet exceeds that of the disk, providing an estimate for the alignment radius ($40\rg$ from Fig.~\ref{fig:angmomflux}b). For model \texttt{T30}, the distance to which the disk aligns to $\mathcal{T}<10^{\circ}$ increases over time, roughly settling at the expected alignment radius of $40\rg$ (see cyan line in Fig.~\ref{fig:tilt_rad}). We also find that the jet power remains roughly constant over the alignment period, suggesting that the alignment process proceeds via angular momentum exchange without any dissipative losses, i.e. elastically. Further, the alignment process will proceed through the disk wind region. This wind-mediated angular momentum exchange would also incur minimal energy losses as the wind region itself is small, being strongly squeezed between the jet and the disk. 

For our extremely misaligned MAD models, the brief jetted periods also show magneto-spin alignment, with one notable difference: the head of the aligned section of the jet collides with the disk resulting in large dissipation, as can be seen from the reduced jet power at large radii in \texttt{T90L} in Fig.~\ref{fig:phibh}. On the other hand, the jet base at small radii, continues to align the disk during the alignment period via the same mechanism discussed above. The jet becomes unstable as it disrupts magnetic flux supply from the disk upon alignment as discussed in the previous section.

We also do not find a direct connection to the so-called MAD saturation radius \citep[$r_{\rm MAD}$; e.g.,][]{narayan03}, i.e., the distance to which the disk is magnetically dominated. The moderately-misaligned MAD simulations in this work show disk alignment out to $100\,r_{\rm g}$, where the disk is at least partially magnetically saturated. It is presently unknown what dictates the MAD saturation radius and how far out can it reach, with current simulations suggesting $r_{\rm MAD} \gtrsim 40-100\,r_{\rm g}$ \citep[][]{narayanSANE2012, Chatterjee:2022}. The MAD saturation radius could possibly become important for alignment if $r_{\rm MAD}\lesssim r_{\rm msa}$, which could occur in truncated disks.

Since the alignment process depends on the jet angular momentum, we expect there to be a strong correlation between alignment and jet power. We check the dependence of $\mathcal{T}_{\rm disk}$, evaluated at $r=10\rg$, on $\eta_{\rm out}$ in Fig.~\ref{fig:tdisk_vs_eta}. For the moderately-misaligned disk models, the MAD state develops quickly and the alignment of the inner disk ($r\approx 10\,r_{\rm g}$) occurs within the first $10^4\,r_{\rm g}/c$ of the simulation (as seen from Fig.~\ref{fig:tilt}). This is why these particular models (black circles) cluster near $\mathcal{T}_{\rm disk}\approx 0^{\circ}$ and $\eta_{\rm out}\gtrsim 100\%$. On the other hand, the extremely misaligned models (red circles) exhibit an incredibly large variation of disk and jet morphologies, spanning almost all possible values of disk tilt angles and jet efficiencies, bounded from above by the relation $\mathcal{T}_{\rm disk, max}\propto 1/\sqrt{\eta_{\rm out}}$.  

We include three additional $30^{\circ}$ tilted MAD disk simulations varying the BH spin direction and magnitude: $a = \pm 1/2$ and $-0.9375$, indicated by cyan circles in Fig.~\ref{fig:tdisk_vs_eta} (including the previously mentioned $\tc{}$ simulation). We provide more details about the simulations in Table~\ref{tab:tiltmod}. These simulations exhibit partial alignment of the inner disk even in the MAD state. Figure~\ref{fig:tdisk_vs_eta} shows that the jet efficiencies from these BH disks are relatively small ($\eta_{\rm out}<60\%$) compared to the $a=+0.9375$, $30^{\circ}$ tilted disk model. Thus the resultant jet torque is not strong enough to align the disk with the BH symmetry plane and the disk tilt angle falls well below the $1/\sqrt{\eta_{\rm out}}$ upper bound. 

Given the above results, we suggest that the jet efficiency upper bound is a robust relation due to the magneto-spin alignment mechanism, independent of BH spin and initial disk misalignment angle. Even under the condition of abundant supply of magnetic flux to the BH, the accretion flow within the inner $10\,r_{\rm g}$ of the BH (and therefore, also the jet base) cannot be misaligned with the BH spin axis by an angle larger than $\mathcal{T}_{\rm disk/jet, max}\approx 70^{\circ}\sqrt{20\%/\eta_{\rm out}}$. We expect such a relation to also hold for BHs accreting at larger accretion rates as $\eta_{\rm out}$ typically decreases for smaller disk scale heights. At sufficiently large accretion rates ($\dot{M}\gtrsim 0.1\%$ of the Eddington accretion rate), the disk cools and becomes geometrically-thin. In such disks, turbulent viscosity plays a more important role in damping the disk warp, and we expect Bardeen-Petterson alignment over magneto-spin alignment \citep{bardeen75, polko17, Liska:19}.

\subsection{Jet power with misaligned disks}
\label{sec:jet_power}

\begin{figure}
    \centering
    \includegraphics[width=\columnwidth,trim= 0 0 0 0, clip]{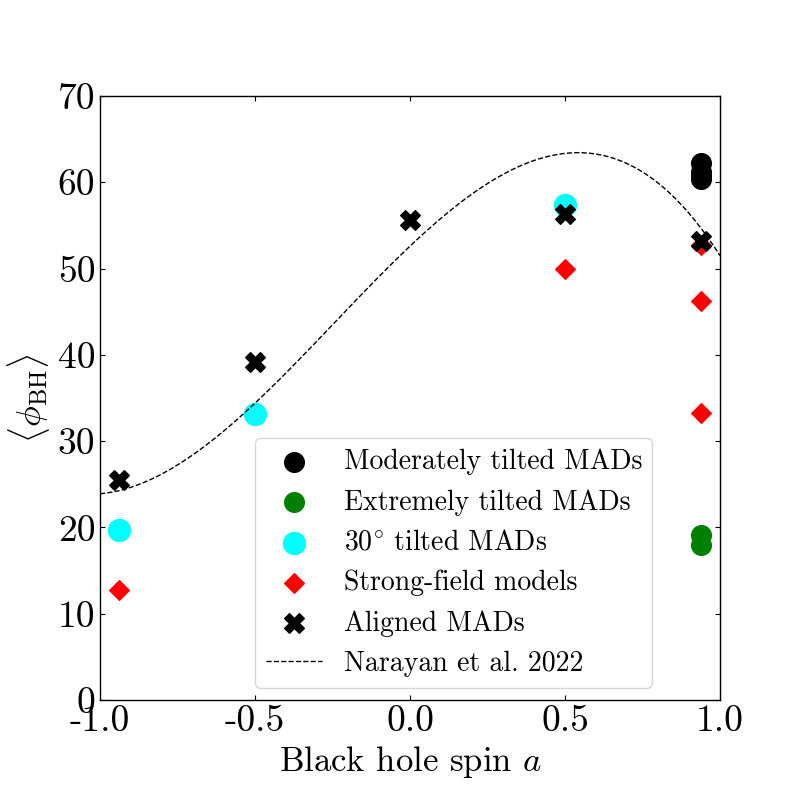}
    \includegraphics[width=\columnwidth,trim= 0 0 0 0, clip]{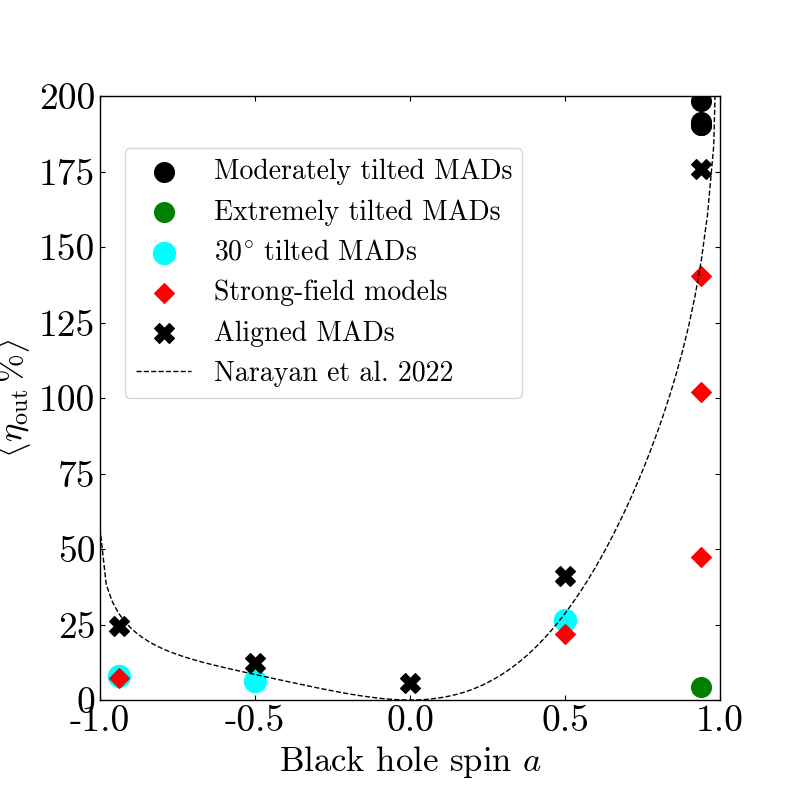}
    \caption{Normalized magnetic flux at the horizon (top) and Outflow efficiency plot (bottom) for all models, time-averaged over the final $10^4\,r_{\rm g}/c$ of each simulation.
    }
    \label{fig:eta}
\end{figure}

One of the most important attributes of the MAD state is the production of powerful jets with efficiencies $\eta_{\rm out}>100\%$ when the BH spin $a\gtrsim 0.8$ \citep{tch11,tch12proc,Narayan:2022,Chatterjee:2023_JP}. The moderately misaligned disk models in our fiducial model set (\tb-\te) exhibited magneto-alignment of the inner disk and reached the MAD state with outflow efficiencies exceeding 100\% (Table~\ref{tab:tiltmod}). The highly misaligned models, on the other hand, showed weakly powered jets ($\eta_{\rm out} \lesssim 5\%$), due to the lack of sustained jet production. Thus, our results suggest that, given enough magnetic flux supply, for initial disk misalignment angles $\lesssim 60^{\circ}$, magneto-alignment of the inner disk due to the emergence of the MAD state would result in jet efficiencies equivalent to the MAD limit for the corresponding aligned disk model. 

However, for misaligned disks that are initialized with strong poloidal magnetic fields but not enough magnetic flux to reach the MAD state (hereafter called ``strong-field'' models), the disk bulk exhibits radial tilt oscillations. Only the inner disk undergoes partial alignment with the BH spin and accretion occurs via plunging streams. To show what happens to $\phi_{\rm BH}$ and $\eta_{\rm out}$ for strong-field misaligned disks, we include 3 misaligned disk models with BH spin $a=0.9375$ from \citet{liska_tilt_2018} and \citet{Chatterjee_2020} - for initial disk misalignment angles of $0^{\circ}$ (\texttt{T0-S}), $30^{\circ}$ (\texttt{T30-S}) and $60^{\circ}$ (\texttt{T60-S}). In addition, we include two previously unpublished simulations from the strong-field model set where the initial disk tilt is $30^{\circ}$ but the BH spin is $a=0.5$ (\texttt{T30A5-S}) and $a=-0.9375$ (\texttt{T30A93M-S}). Table~\ref{tab:tiltmod} shows a summary of the simulation parameters. Further details about the simulations are given in \citet{liska_tilt_2018} and \citet{Chatterjee_2020}. For comparison, we also included a series of aligned MAD simulations with BH spins $a=0,\pm0.5,\pm0.94$ from the \hammer{} EHT simulation library \citep{EHT_SgrA_2022_PaperV} as well as the fits for $\phi_{\rm BH}$ and $\eta_{\rm out}$ in the aligned MADs from \citet{Narayan:2022}. 

Figure~\ref{fig:eta} shows the normalized magnetic flux at the event horizon, $\phi_{\rm BH}$, and the outflow power efficiency, $\eta_{\rm out}$, for both the MAD and the strong-field models. First, as expected, the moderately-misaligned MAD models show similar magnetic flux and jet power values as the aligned models while the extremely-misaligned models are flux-depleted and therefore, host very weak jets. Second, for the $a=0.9375$ strong field models, the magnetic flux (and thus, the jet power as well) decreases significantly with increasing initial disk angle. In these models, the warping of the accretion flow prevents efficient accumulation of magnetic flux near the BH (i.e., the warped disk is unable to ``hold on'' to the magnetic flux as effectively as in aligned disks). 

Next, for low-spin BHs ($a=0, \pm 1/2$ in our model set), we see little-to-no difference between the aligned and misaligned disk $\phi_{\rm BH}$ and $\eta_{\rm out}$ values as the Lense-Thirring torque is either too small or absent. Additionally the resultant jets are too weak to force the disk into alignment (which also supports the $\mathcal{T}_{\rm disk, max} \propto 1/\sqrt{\eta_{\rm out}}$ relation introduced in the previous subsection). Finally, for rapidly spinning retrograde BHs, we again see that disk warping lowers the magnetic flux value compared to aligned MADs, producing weak jets that are unable to magneto-align the disk as $\eta_{\rm out} \approx 10\%$. Overall, we find that introducing misaligned accretion lowers the magnetic flux and the jet power in high spin (prograde or retrograde) BHs as Lense-Thirring torques are large enough to warp the disk with the exception of high spin prograde MADs when the jet can become strong enough to magneto-spin-align the disk.

\section{Astrophysical implications}
\label{sec:discussion}

\subsection{Quasi-periodicity in AGN observations due to jet-disk collisions}
\begin{figure*}
    \centering
    \includegraphics[width=0.32\textwidth,trim= 8cm 0 0cm 4cm, clip]{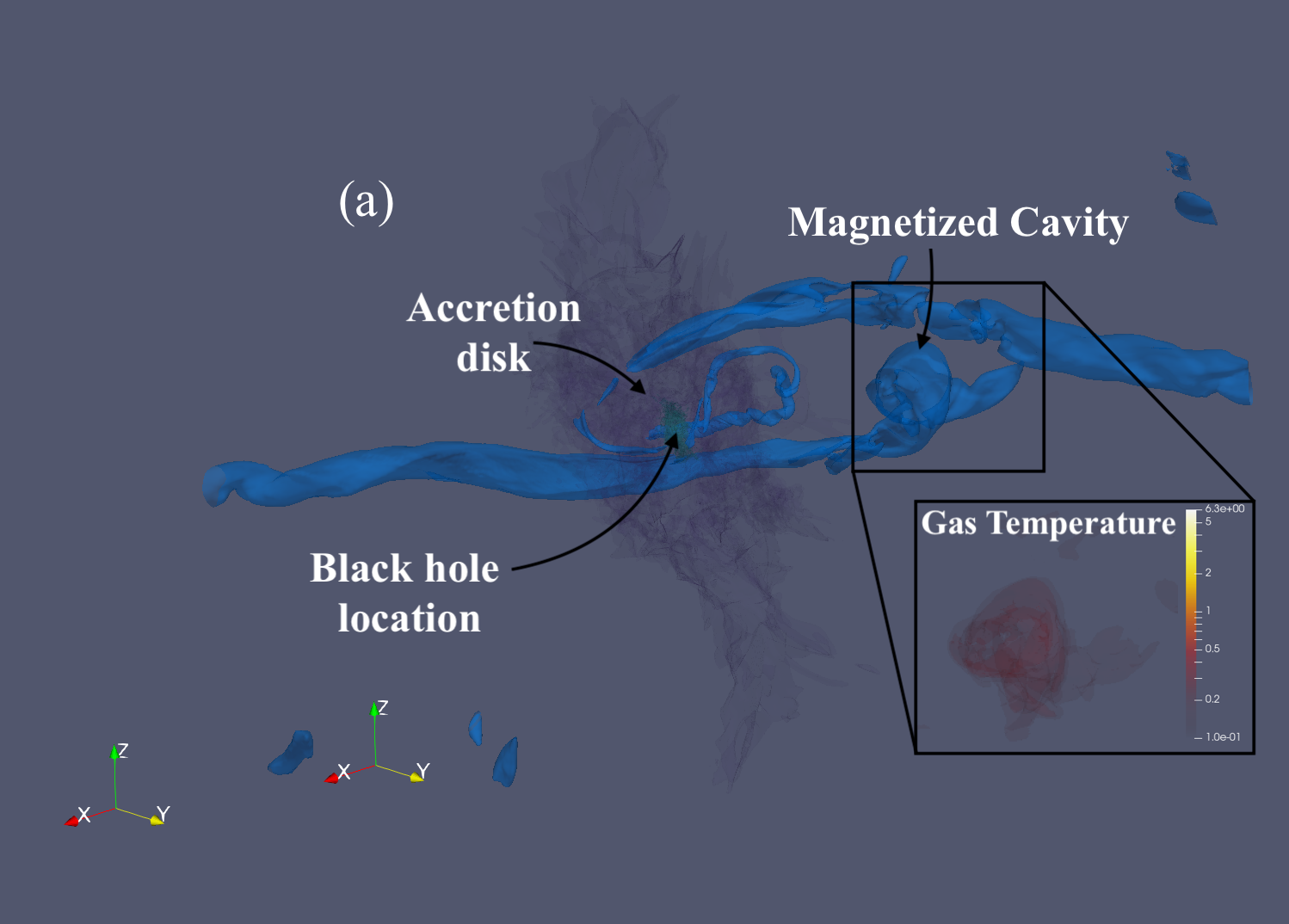}
    \includegraphics[width=0.32\textwidth,trim= 8cm 0 0cm 4cm, clip]{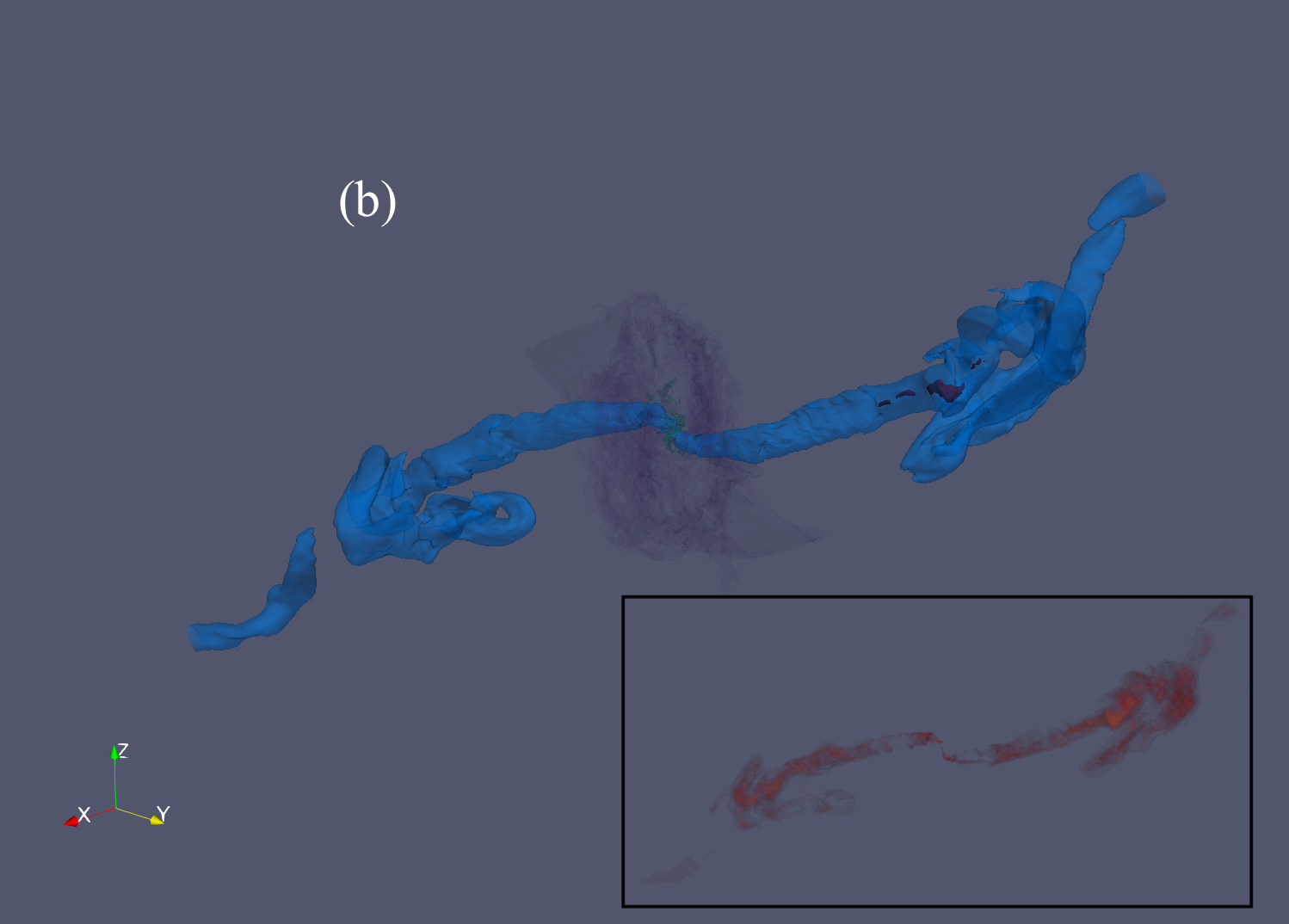}
    \includegraphics[width=0.32\textwidth,trim= 8cm 0 0cm 4cm, clip]{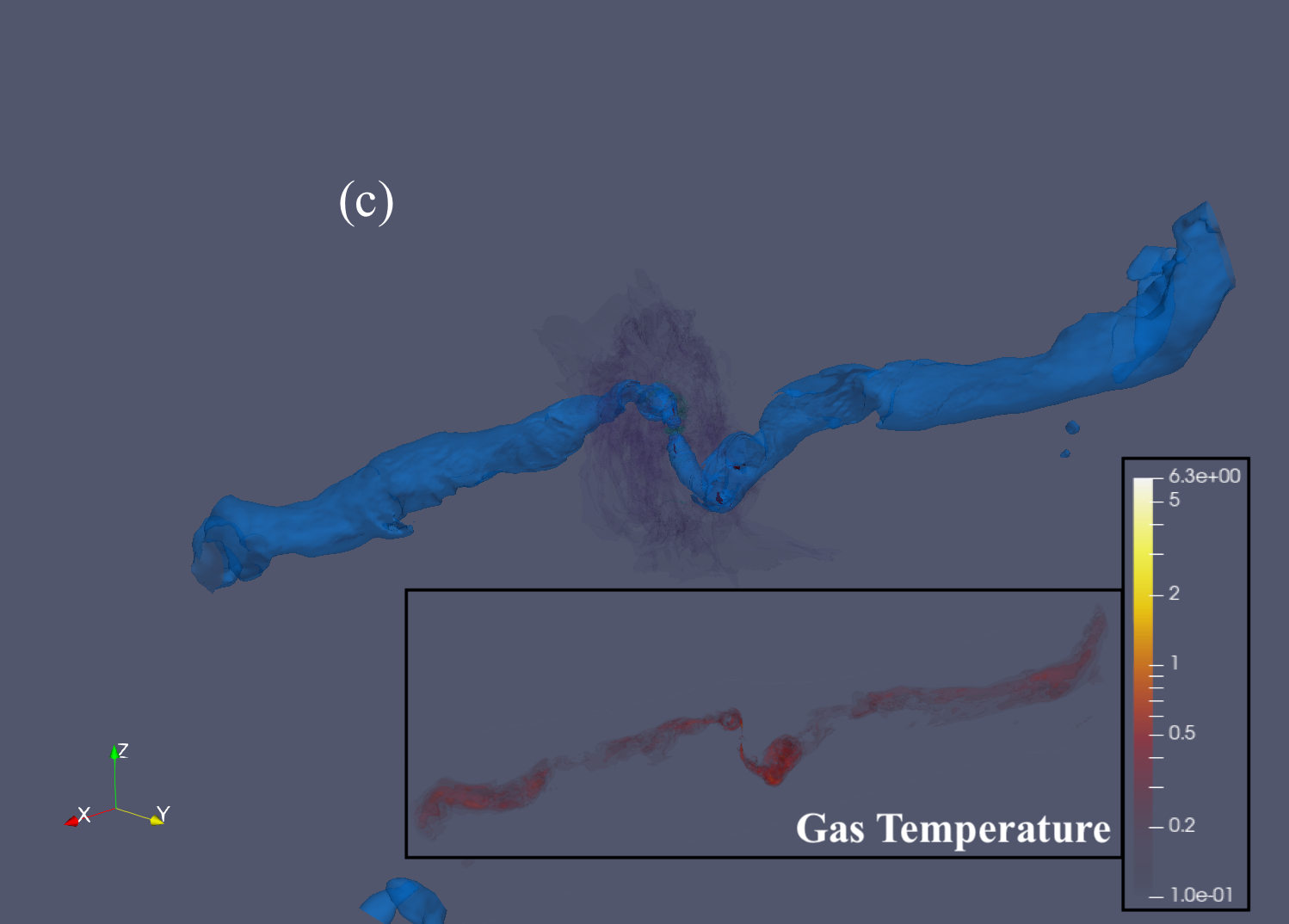}
    \includegraphics[width=0.32\textwidth,trim= 8cm 0 0cm 4cm, clip]{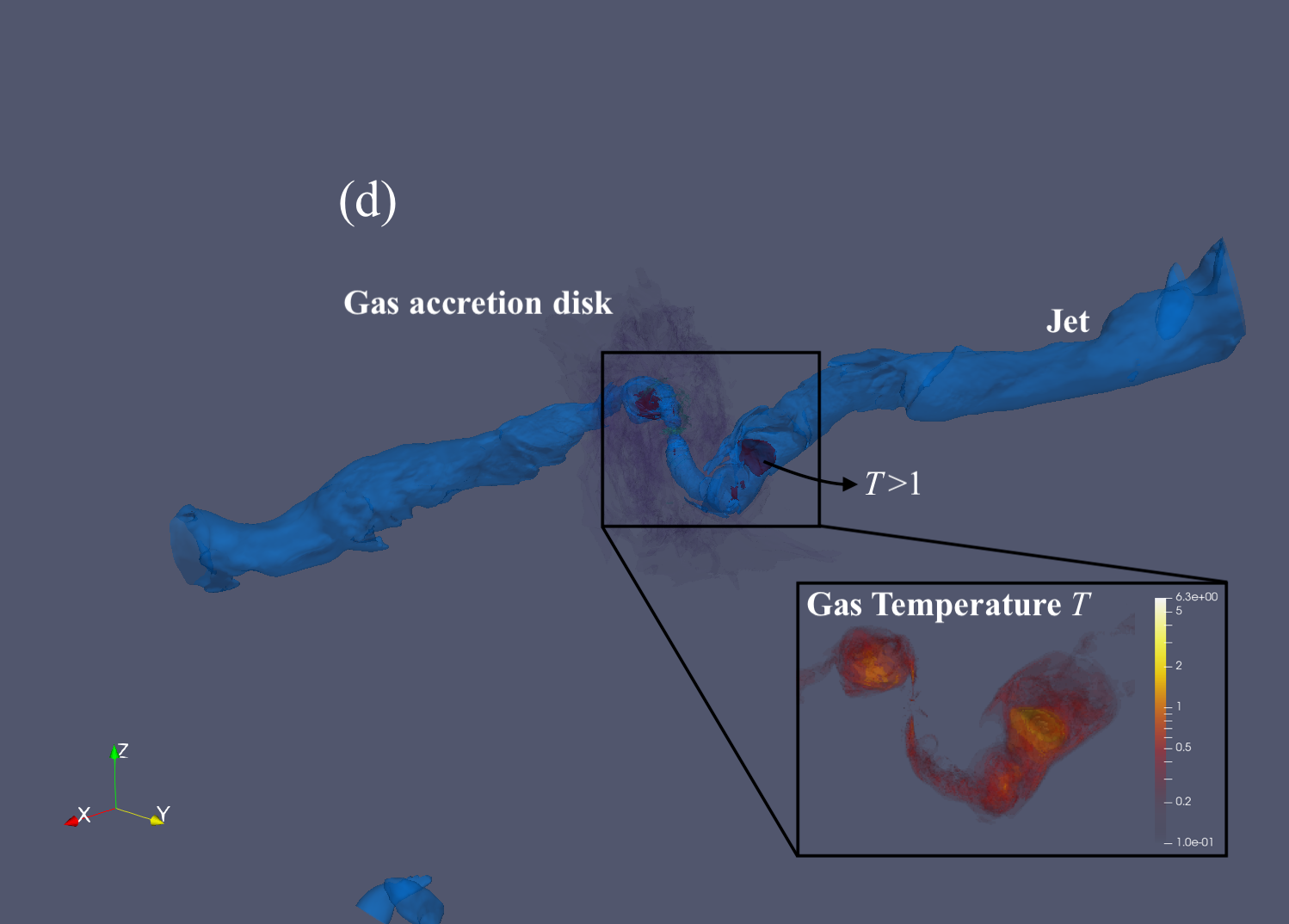}
    \includegraphics[width=0.32\textwidth,trim= 8cm 0 0cm 4cm, clip]{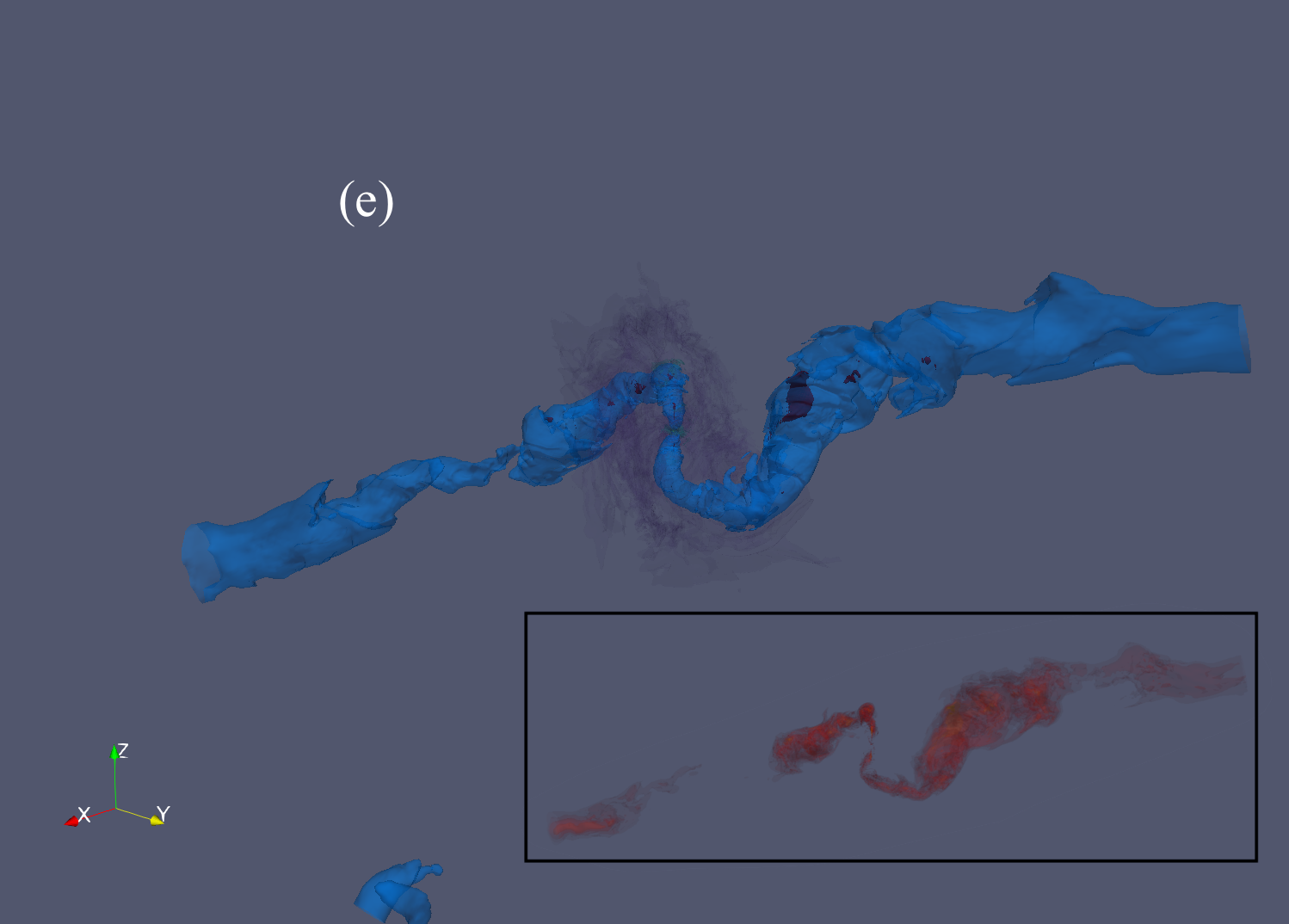}
    \includegraphics[width=0.32\textwidth,trim= 8cm 0 0cm 4cm, clip]{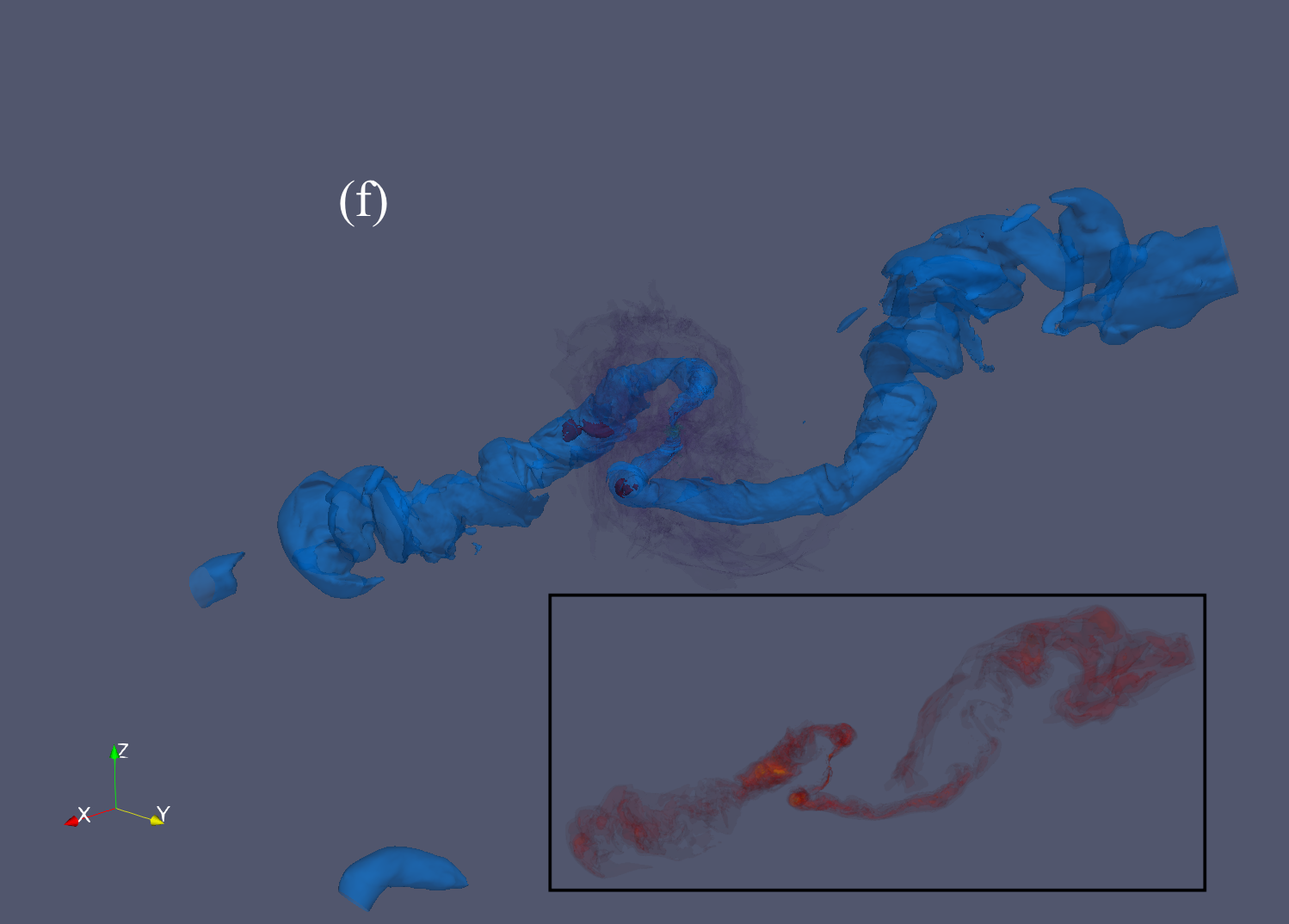}
    \includegraphics[width=0.32\textwidth,trim= 8cm 0 0cm 4cm, clip]{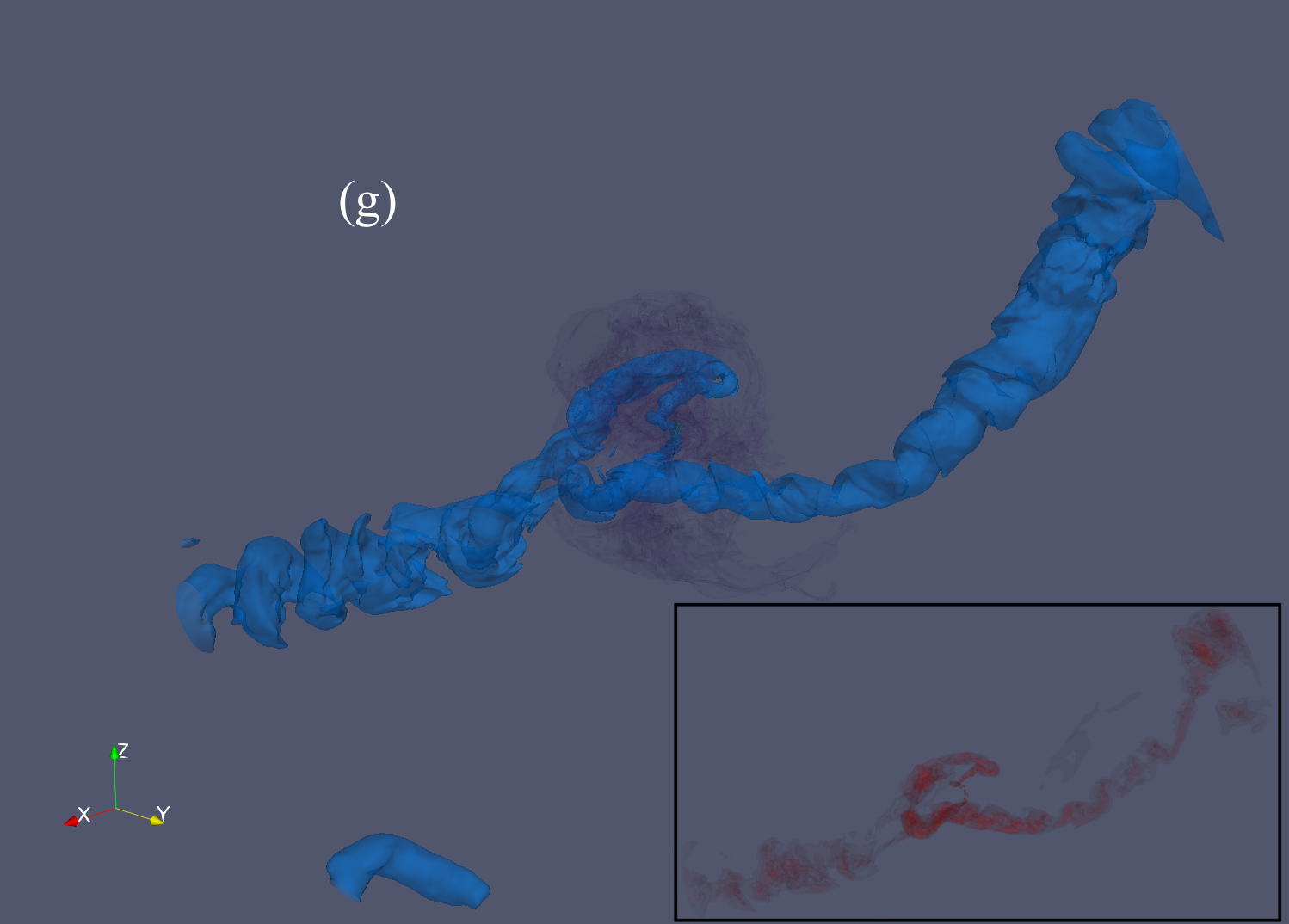}
    \includegraphics[width=0.32\textwidth,trim= 8cm 0 0cm 4cm, clip]{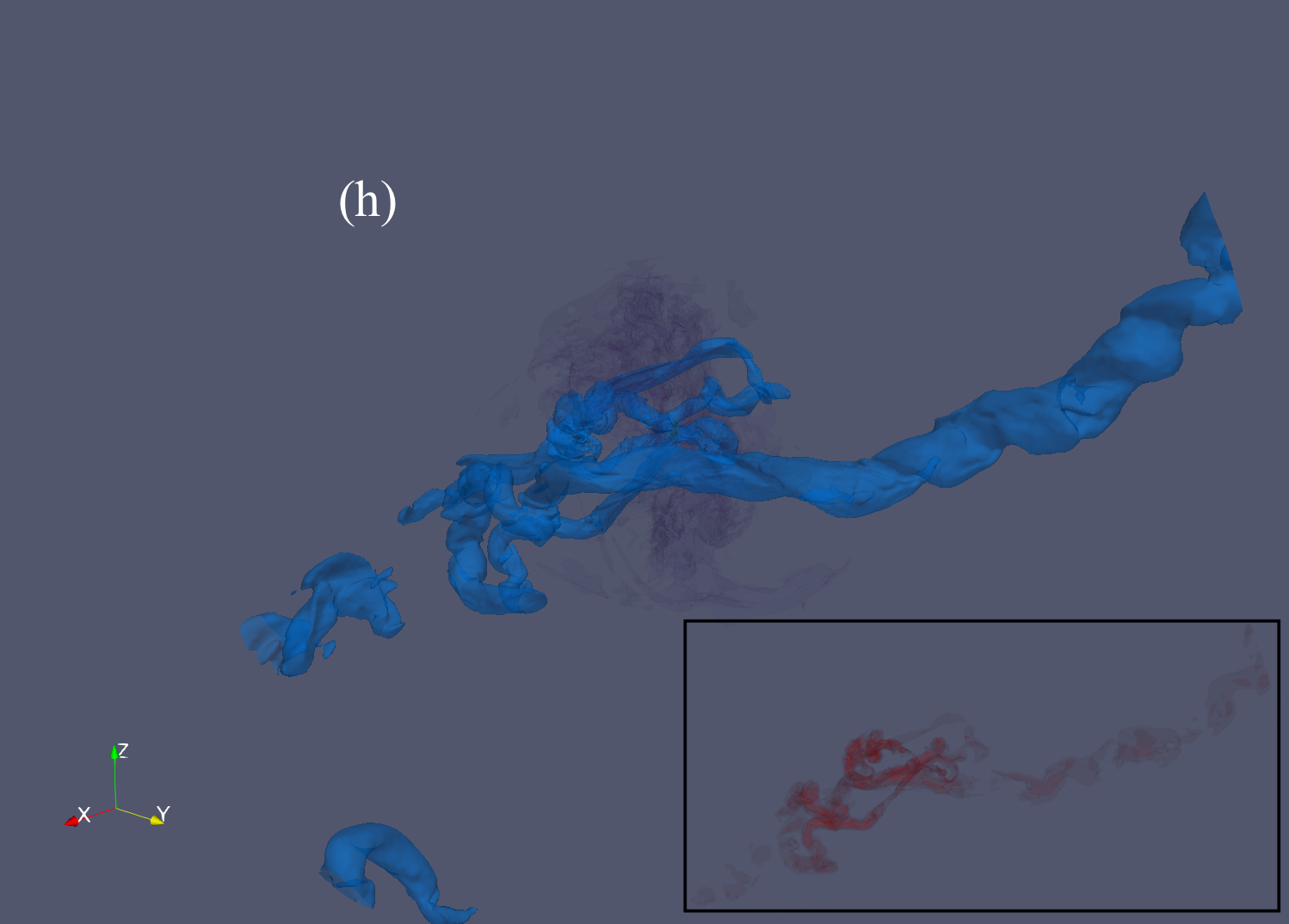}
    \includegraphics[width=0.32\textwidth,trim= 8cm 0 0cm 4cm, clip]{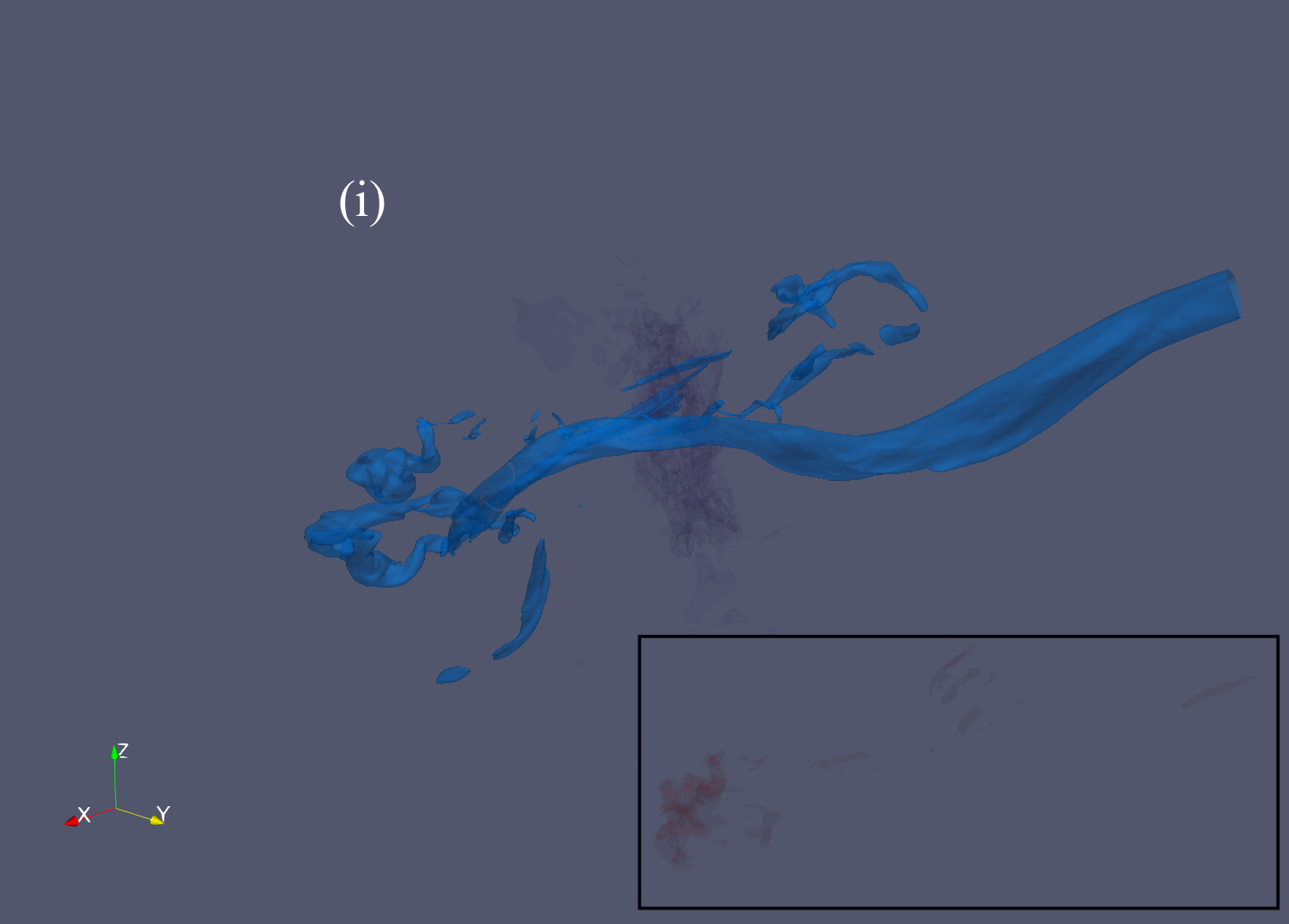}
    \caption{3D renderings of \ti{} during a jet ejection phase: the disk (indicated by gas density in green-purple) and the jet (indicated by the magnetization $b^2/\rho c^2=1$ surface in blue). The insets show gas temperature in relativistic units ($\Theta \equiv k_{\rm B}T_{\rm gas}/m_{\rm p}c^2\geq1$ indicates relativistically-hot gas). The box size is the same as the simulation box (i.e., a sphere with a maximum radius of $10^3\,\rg{}$). An animation is available on \href{https://youtu.be/mTLbZgd5CtY}{YouTube}. }
    \label{fig:cavities}
\end{figure*}

Our simulations reveal two exciting regimes of misaligned accretion, (1) one where the torque due to the jet dominates over the accretion flow, forcing the inner disk into alignment with the BH spin axis at least within a few tens of gravitational radii from the BH, and (2) the other where jet ejection is intermittent in nature and leads to a rapid re-orientation of the accretion flow. While the first regime, seen for MAD disks with $\mathcal{T}_{\rm disk}\lesssim60^{\circ}$, resembles aligned MAD accretion, the extreme misaligned disks with $\mathcal{T}_{\rm disk}\gtrsim70^{\circ}$ exhibit a variety of disk/jet morphologies, especially during outbursts. 

Figure~\ref{fig:cavities} shows the evolution of one such outburst phase seen in the high resolution model \ti{}, with the initial disk tilt, $\mathcal{T}_{\rm disk}=90^{\circ}$. In Fig.~\ref{fig:cavities}(a), we start with a snapshot in time when the jet is inactive, but the environment is filled with filamentary magnetized remnants (in blue) from a previous outburst. These remnants, at times, are also filled with near-relativistically hot plasma (with $\Theta\equiv k_{\rm B}T_{\rm gas}/mc^2\gtrsim 0.1$) and buoyantly rise outwards, resembling hot cavities often seen in AGN \citep[e.g.,][]{Ubertosi:2021}. Once enough magnetic flux accumulates at the BH event horizon, the jet relaunches. This can be seen in Fig.~\ref{fig:cavities}(b) where the bipolar structure of the restarted jet is clearly seen pushing out the disk and pre-existing jet remnants. There are also hints of short-lived jet precession from the clockwise movement of the jet base during this early jet ejection phase.

Over time, as the BH horizon magnetic flux increases, the jet grows in strength and tends to align with the vertical BH spin axis, thus forcing the disk into alignment, launching the magneto-spin alignment process. Figure~\ref{fig:cavities}(c,d) shows the jets, as they punch through the disk leaving the large-scale jet in a completely different orientation compared to the small-scale jet. The jet-disk collisions result in regions of super-heated shocked gas (of relativistic gas temperatures $\Theta>1$), potentially being zones of hard X-ray and possibly even $\gamma-$ray emission. 

As we saw in Sec.\ref{sec:extreme}, in colliding with the accreting gas, the jet disrupts its own supply of gas and magnetic flux and triggers a series of severe kink instabilities that cause the jet to wobble vigorously (Fig.~\ref{fig:cavities}e,f) and eventually disrupt (Fig.~\ref{fig:cavities}g,h). Thus the system returns to the state it started with: quasi-spherical weakly-magnetized accreting gas with disordered magnetized filaments and plumes of hot gas, remnants of a self-destructed jet (Fig.~\ref{fig:cavities}i). 

Our larger scale simulation, $\tj{}$, showed that the gas and magnetic flux, ejected during the jet outburst ($2.5\times10^4\,r_{\rm g}/c\lesssim t\lesssim4\times10^4\,r_{\rm g}/c$), re-accumulates near the BH and restarts the outburst at $t\sim 9\times10^4\,r_{\rm g}/c$, or around 12 days later for Sagittarius A$^*$-like BHs and 50 years for M87-like BHs, with each outburst leading to pronounced X-ray flares. The self-consistent nature of the feedback mechanism could also exhibit quasi-periodicity in radio and X-ray eruptions, such as the recently discovered quasi-periodic eruption event of Swift J0230+28 \citep{Evans:2023,Guolo:2023}. For this particular source, the estimated BH mass between $1.5-10$ million $M_{\odot}$, which gives us an approximate periodicity of 4.3 to 28.6 days, consistent with the observed 22 day periodicity. We leave the necessary radiative transfer calculations for detailed comparisons to future work.

Other than these outbursts, separated by days to several years depending on the BH mass, steady jets, if present, are weakly powered ($\eta_{\rm out}\lesssim 5\%$) and may not provide any substantial radio emission, making such highly-misaligned systems good candidates for radio-quiet sub-Eddington AGN. The BZ jet power efficiency is given by $\eta_{\rm BZ}\propto a^2 \phi_{\rm BH}^2$. Thus, to explain the absence of a steady-powered, radio jet, either the BH spin magnitude must be small \citep[as suggested by e.g.,][]{tch10a} or the magnetic flux supply to the BH must be meager. While most BH spin measurements skew towards high spin values \citep{Reynolds:2021}, BHs in galaxies with significant merger activity may carry low spins. For the magnetic flux, the ISM supplies strong magnetic fields \citep[plasma-$\beta\approx 1$, e.g., ][]{Ferriere:2020}, providing an environment favorable for the development of the MAD state. If the BH and disk angular momentum vectors are misaligned enough to allow for jet-disk collisions, both spin and magnetic flux constraints are unnecessary. Capturing radio-quiet AGN during outbursts would then be crucial to verifying whether such highly-misaligned BH systems are indeed viable in nature.

\subsection{Jet precession in highly sub-Eddington black holes}
\label{sec:no_jetprec}

In this section, we examine Lense-Thirring (LT) precession in the context of black holes (BHs) accreting at highly sub-Eddington rates, $\dot{M}\ll 10^{-3}\dot{M}_{\rm Edd}$, where the flow remains geometrically thick and advection-dominated out to large radii ($r\gtrsim100-1000\rg$). This regime corresponds to the quiescent and hard states in BH XRBs and low-luminosity AGN, prior to transitioning into a truncated disk state \citep[e.g.,][]{fend04a,Liska:2024}. GRMHD simulations of this accretion regime, including those in this work, are typically initialized with a rotating torus of gas with a disk height-to-radius ratio, $h/r\approx 0.3-0.6$, and a size $r_{\rm disk}\sim 50-1000\,r_{\rm g}$, depending on the required magnetic flux. Weakly magnetized disks are often initialized with compact tori \citep[$\sim50\,r_{\rm g}$; e.g.,][]{Gammie:03, Porth:19}, while MAD configurations require large-scale poloidal field loops and hence, tori extending to $1000\,\rg$ \citep[e.g.,][]{Tchekhovskoy:2015, Chatterjee:2022}. 

These tori evolve through MHD instabilities, launching outflows and spreading viscously into geometrically thick disks over time ($h/r\gtrsim0.3$). Simulations in this work, along with those of \citet{liska_tilt_2018} and \citet{Chatterjee_2020}, follow similar evolution but with misaligned accretion disks. Initially, a misaligned disk behaves like a quasi-rigid body and undergoes LT precession before expanding radially and settling into a non-precessing warped disk (as seen in Fig.~1e,f in \citealt{liska_tilt_2018} and Fig.~2e in \citealt{Chatterjee_2020}), consistent with the bending wave solution \citep{ivanov97}. This is also why it is important to evolve tilted disks to runtimes long enough to develop a quasi-steady-state disk. We generally find simulation runtimes of $t\gtrsim 10^5\,r_{\rm g}/c$ to be enough for this purpose. The viscous expansion of the disk is crucial for achieving the quasi-steady-state warped disk as under-resolved disks continue to behave as rigid bodies and carry on precessing \citep[see Fig.~1e of][for resolution comparisons]{liska_tilt_2018}. If the numerical resolution is insufficient to resolve the MRI in the outer disk, the disk will not be able to viscously expand, and hence precess at a constant rate. 

Disks develop a steady warped morphology when LT torques are significant (i.e., for high BH spin). However, as we show in this work, misaligned MAD flows around high-spin prograde BHs are an exception: jet torques align the inner disk, suppressing both disk and jet precession. Overall we suggest that jets from resolved misaligned, geometrically-thick accretion disks \textit{do not} exhibit LT precession \citep[also see][]{Nixon:2013}, implying that QPOs seen in the low hard state of BH XRBs may not be associated with the LT mechanism \citep{Fragile:2023}. While this conclusion applies to highly sub-Eddington flows where the disk remains thick out to large radii, there is encouraging evidence from recent simulations that a magnetized, thick inner disk fed by an outer geometrically-thin disk, i.e., a truncated disk, could exhibit precession if the inner thick disk remains compact and rigid \citep{Bollimpalli:2023, Bollimpalli:2024}, a scenario likely at $\mdot\gtrsim 10^{-3}\mdot_{\rm Edd}$ \citep{Liska:2024}. Disk precession has also been demonstrated for highly misaligned (initial $\mathcal{T}_{\rm disk}=65^{\circ}$), thin ($h/r\approx0.03$) disks, where the inner disk tears off and precesses independently over multiple orbits \citep{Nixon:2012,liska:2021_tearing, liska_hamr:2022,liska_coronal_lag_2023}. Disk tearing could be a viable candidate for explaining X-ray QPOs seen in the hard-intermediate state of BHXRBs \citep{Musoke:2023}.

\begin{figure}
    \centering
    \includegraphics[width=\columnwidth,trim= 0 0 0 0, clip]{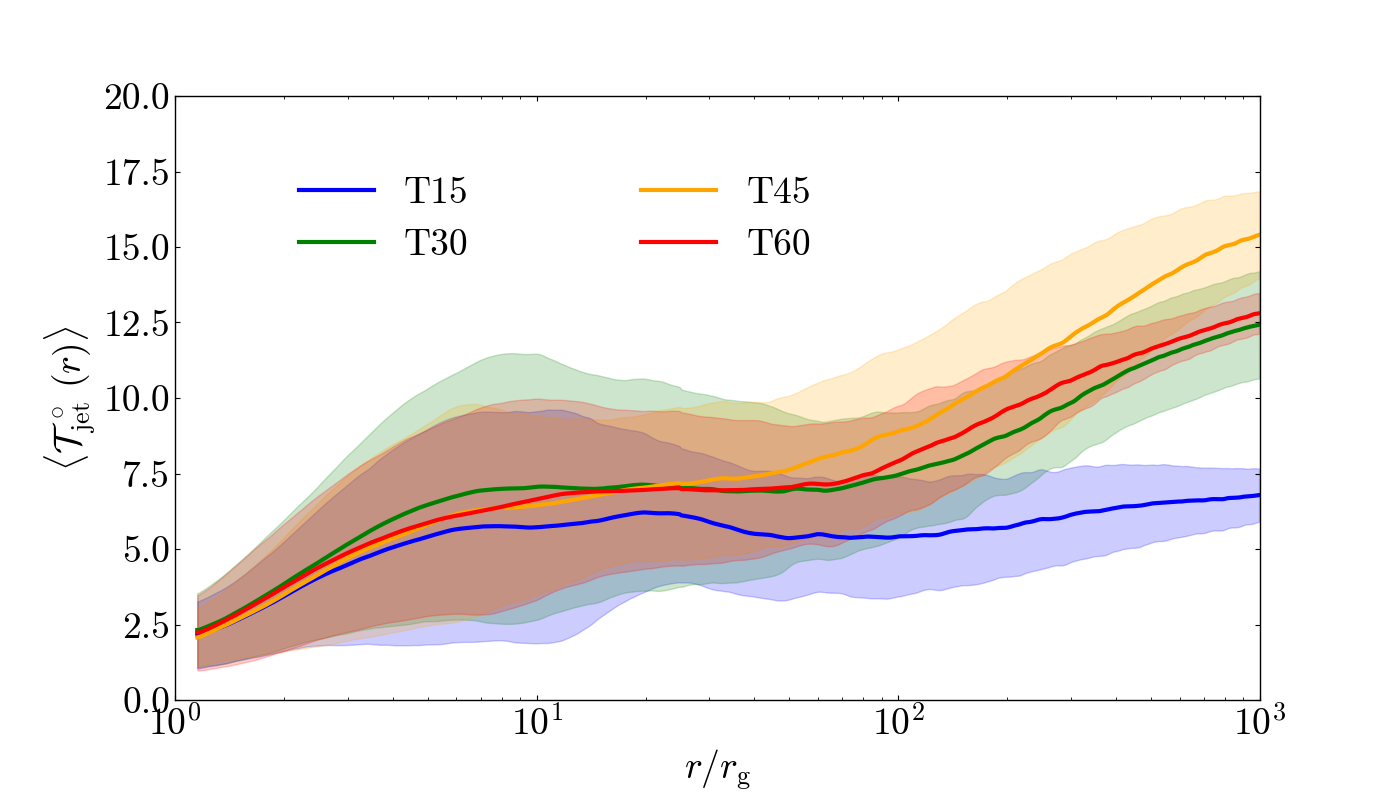}
    \includegraphics[width=\columnwidth,trim= 0 0 0 0, clip]{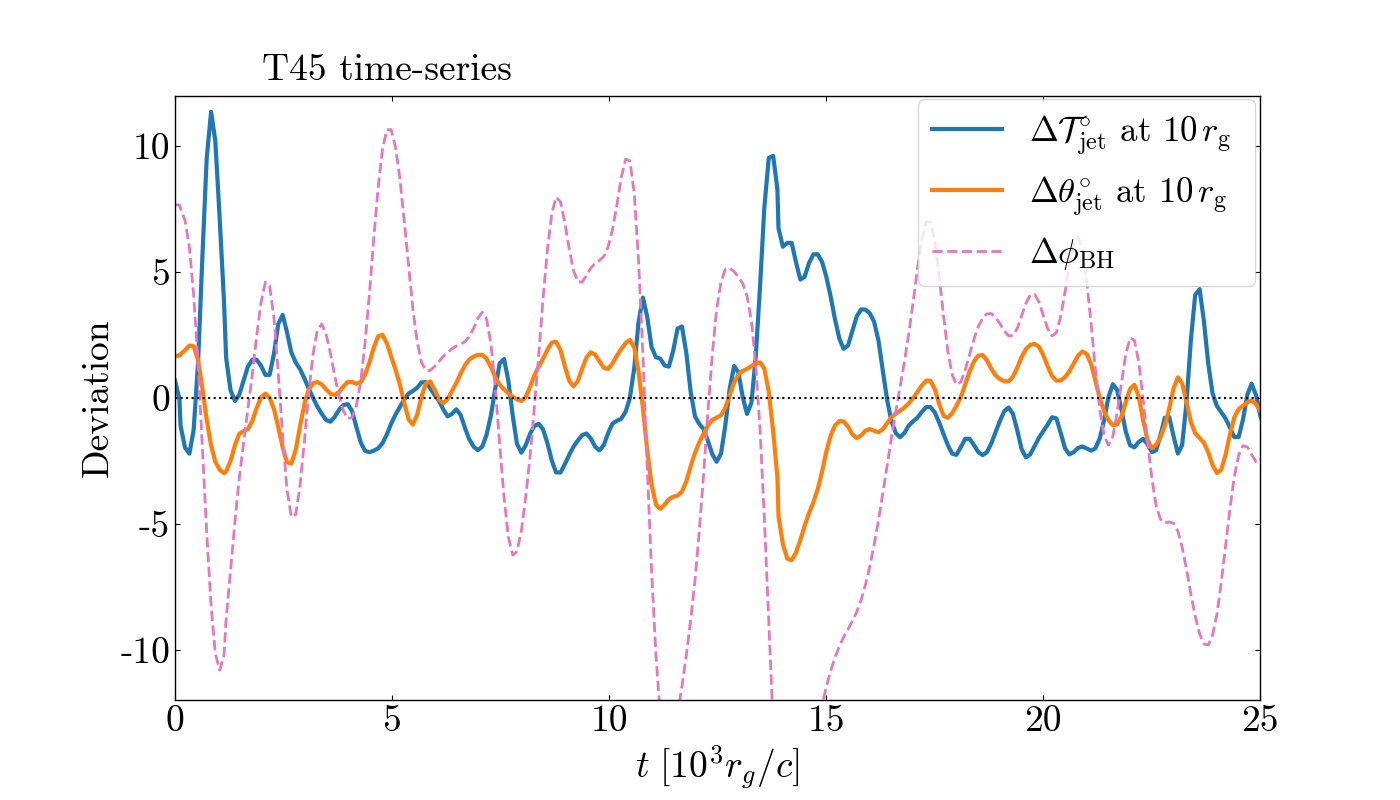}
    \caption{Despite alignment, jets wobble significantly due to magnetic flux eruptions and can appear to be precessing over periods of several years (for a $10^9\,M_{\odot}$ BH). (Top) We show radial profiles of the jet tilt angle with 1$\sigma$ standard deviation for the moderately misaligned MAD models \texttt{T15-60}. (Bottom) We find that deviations in the jet opening angle $\theta_{\rm jet}$ and orientation $\mathcal{T}_{\rm jet}$ are correlated and anti-correlated with the horizon magnetic flux $\phi_{\rm BH}$. We track time-series curves of the deviations of the horizon magnetic flux and the jet tilt angle from their respective mean values for one model \td{} over $5\times10^4\,r_{\rm g}/c$. The jet becomes wider ($\theta_{\rm jet}$ increases) as the MAD state develops ($\phi_{\rm BH}$ increases). Large excursions in $\mathcal{T}_{\rm jet}$ align with drops in $\phi_{\rm BH}$ caused by magnetic flux eruption events.}
    \label{fig:jettilt}
\end{figure}

\subsection{Periodicity in jet observations}
\label{sec:jet_obs}

Lightcurve periodicities and precession-like wobbling are also seen in AGN jets: possible models involve orbital precession due to candidate binary black holes \citep[e.g., for OJ287,][]{Britzen:2023} and LT precession. Further, the jet in M87 has a measurable transverse displacement with a quasi-periodicity of 8-10 years \citep{walker_2018_M87}. If not LT precession, fluid instabilities such as the Kelvin-Helmholtz instability (KHI) are invoked to partially explain jet wobbling and apparent precession, but there are also counter-arguments to this reasoning, chief of which is whether KHI can actually drive large-scale lateral motion. Other than KHI, collisions between the jet base and the disk wind can also drive oscillations in the jet orientation. Jets from MAD accretion flows (aligned or tilted) exhibit wobbling with $1\sigma$ standard deviation of $3-5^{\circ}$ in the jet tilt angle (Fig.~\ref{fig:jettilt} and Table~\ref{tab:tiltmod}). Displacements of even a few degrees may become substantial when the jet is viewed at certain inclination angles due to Doppler effects. Such oscillations in the jet base could then potentially explain quasi-periodic shifts in the radio core position of blazar jets \citep[e.g.,][]{Weaver:2022}.

Other than jet-wind collisions, magnetic flux eruption events in MAD could lead to substantial lateral displacements in the jet base. Figure.~\ref{fig:jettilt} (bottom) shows the deviation in jet tilt angle, $\mathcal{T}_{\rm jet}$, (or orientation angle) and the jet opening angle, $\theta_{\rm jet}$, at $10\,\rg$ and the normalized magnetic flux at the BH horizon, $\phi_{\rm BH}$, from their mean values over a period of $5\times10^4\,\rg/c$ (or $\approx 50$ years for M87). Eruption events, indicated by a sharp drop in $\phi_{\rm BH}$, release cylindrical tubes of vertical magnetic fields that then interact with the accreting gas \citep[e.g.,][]{Chatterjee:2022}, and can generate high energy flares \citep{Chatterjee:2021,Ripperda2022}. 

Additionally, during eruption periods, while $\theta_{\rm jet}$ at $10\rg$ drops as well, there are no significantly changes at larger scales During the largest eruption events, $\mathcal{T}_{\rm jet}$ at $10\,\rg$ can change by nearly $10^{\circ}$, and the largest deviations in the jet orientation are spaced roughly $10^4\,\rg/c$ apart, approximately 10 years for M87, potentially providing an alternate origin for precession-like behavior in the jet base \citep[see also][]{Lalakos:2023}. Further, these lateral excursions drive waves through the jet-wind shear layer \citep{Davelaar:2023} that could develop into the large-scale flow patterns often seen in radio images of jets, manifesting in quasi-periodic position angle shifts \citep[e.g., as seen in][]{Cui:2023}. Simultaneous detections of bright flares (from X-rays to TeV energies) with observed lateral shifts in radio jets could verify whether magnetic flux eruptions could indeed be a viable alternative to orbital or LT precession for quasi-periodic jet motion in the sub-Eddington accretion regime.
 
Finally, we note that the time spacing between large magnetic eruption events can vary significantly and could be dependent on the geometrical thickness of the disk. Though we neglect the effects of radiative cooling, our simulations should be applicable for M87 ($\dot{M}\approx 10^{-4}\dot{M}_{\rm Edd}$) as the disk does not lose significant vertical pressure support due to radiative cooling and remains geometrically-thick \citep{Chatterjee:2023, Liska:2024}. We leave the exploration of the properties of magnetic flux eruptions for different misaligned disks to future work.

\section{Conclusions}
\label{sec:summary}

In this work, our goal is to deconstruct the evolution of highly magnetized accretion disks (``MADs'') whose angular momentum axis is misaligned relative to the black hole (BH) spin axis. For this, we simulate an extensive set of GRMHD simulations of disks with 7 different initial tilt angles, ranging from $0^{\circ}$ and $90^{\circ}$, around a Kerr BH of spin parameter $a=0.9375$. Additionally, we spot-check other spin values ($a=-0.9375, \pm 0.5$) and consider other near-MAD magnetic setups. Table~\ref{tab:tiltmod} provides a summary of all our models.

We summarize our results here:
\begin{enumerate}
    \item Given enough magnetic flux supply and a rapidly spinning BH, geometrically-thick disks ($h/r\sim0.3$) tilted at $\mathcal{T}\lesssim60^{\circ}$ can be forced into magneto-spin alignment with the BH spin axis, out to $r\lesssim100\rg$. 
    \item Extremely misaligned MAD disks, $\mathcal{T}\gtrsim75^{\circ}$, around rapidly spinning BHs exhibit intermittent jetted outbursts, separated by quiet periods of $\Delta t \sim 5\times10^4\,r_{\rm g}/c$. Interestingly, in the process of aligning the disk, these jets expel the disk's magnetic flux out to $1000$s of $r_{\rm g}$, inadvertently choking their own flux supply. However, over time, some portion of the disk poloidal flux reaches the BH horizon again, reactivates the MAD state and the jets, and the cycle repeats again. This is the first demonstration of restarted jets in GRMHD simulations and could explain transient radio and X-ray activity (due to shocks via jet-disk interactions) in otherwise low-luminosity BH systems.
    \item Magneto-spin alignment of the disk depends on both the magnitude of the magnetic flux on the BH horizon and the BH spin. Our models suggest that only highly spinning prograde BHs in the MAD state can produce jets powerful enough to magnetically align geometrically-thick accretion disks.
    \item For jets that are unable to magneto-spin-align a tilted disk, the disk shows radial tilt oscillations. In this case, increasing the initial disk tilt angle results in lower horizon magnetic flux values and lower jet powers. The jet power also changes with both the spin and tilt angle as the MAD magnetic flux saturation value is different for different spins, similar to their aligned version \citep[e.g.,][]{tch12proc,Tchekhovskoy:2015,Narayan:2022,Chatterjee:2023_JP,Lowell:2023}.
    \item Finally, combining the results from this work and previous literature for tilted thick disks, we find that the disk/jet does not undergo Lense-Thirring precession for highly sub-Eddington BH accretion flows ($\dot{M}\ll 10^{-3}\dot{M}_{\rm Edd}$). We suggest that the large jet wobbling amplitude seen during magnetic flux eruption events could provide an alternate mechanism for precession-like behavior in AGN jets \citep[e.g., for M87;][]{walker_2018_M87,Cui:2023}, with an approximate time period of several years for a $10^9M_{\odot}$ BH.
\end{enumerate}

\section*{\label{sec:Acknowledgments}Acknowledgments}
We thank the anonymous referee for their careful reading of the text and their suggestions. We thank Ramesh Narayan, Yuri Kovalev and Vladislav Makeev for useful discussions on accretion and curved jets. KC was supported in part by grants from the Gordon and Betty Moore Foundation and the John Templeton Foundation to the Black Hole Initiative at Harvard University, and by NSF award OISE-1743747. ML was supported by the John Harvard, ITC and NASA Hubble Fellowship Program fellowships. AT acknowledges support from the NSF AST-2009884 and NASA 80NSSC21K1746 grants. AT was supported by BSF grant 2020747 and NSF grants AST-2107839, AST-1815304, AST-1911080, AST-2206471, OAC-2031997.This research was enabled by support provided by a INCITE program award PHY129, using resources from the Oak Ridge Leadership Computing Facility, Summit, which is a US Department of Energy office of Science User Facility supported under contract DE-AC05- 00OR22725, as well as Calcul Quebec (http://www.calculquebec.ca) and Compute Canada (http://www.computecanada.ca). This work has made use of NASA's Astrophysics Data System (ADS). This research was supported in part by grant no. NSF PHY-2309135 to the Kavli Institute for Theoretical Physics (KITP).\\

\appendix
\section*{\label{sec:Appendix}Appendix}

\begin{figure*}
    \centering
    \includegraphics[width=\textwidth,trim= 0 0 0 0, clip]{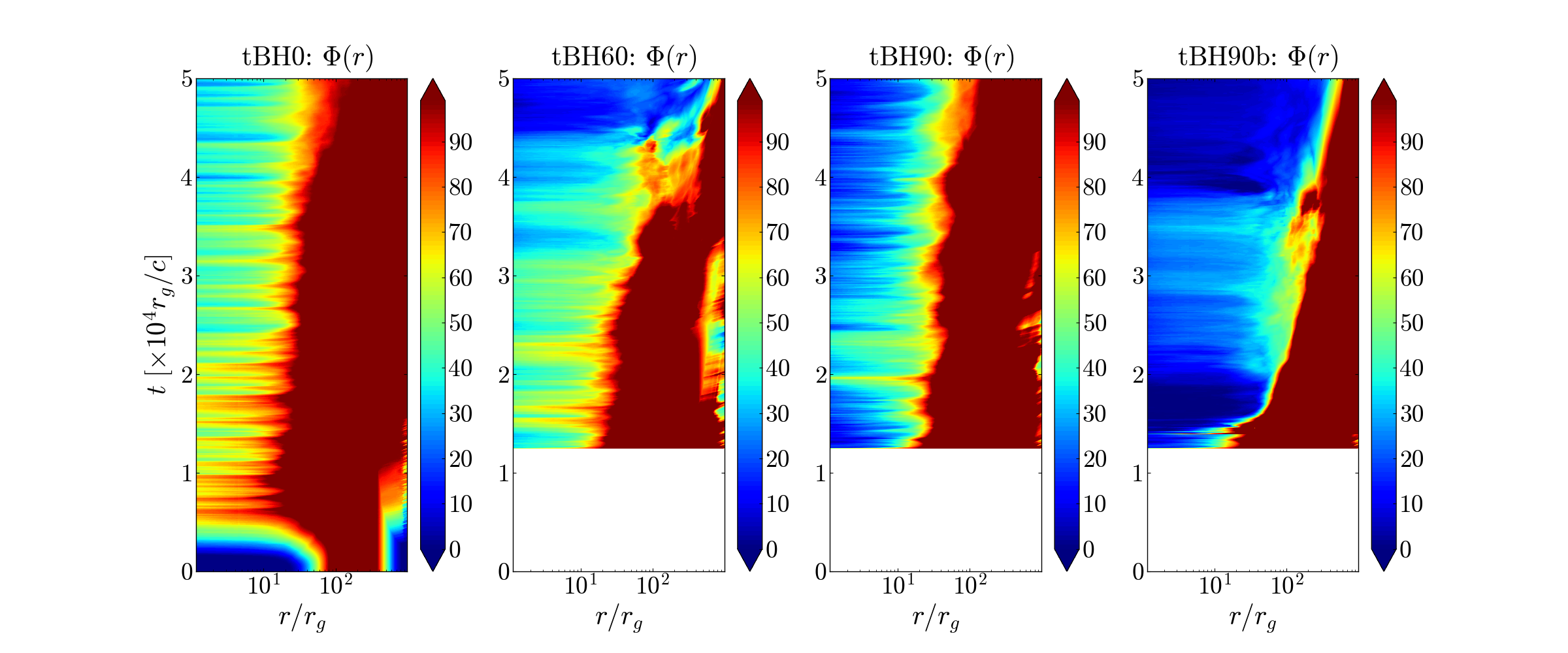}
    \includegraphics[width=\textwidth,trim= 0 0 0 0, clip]{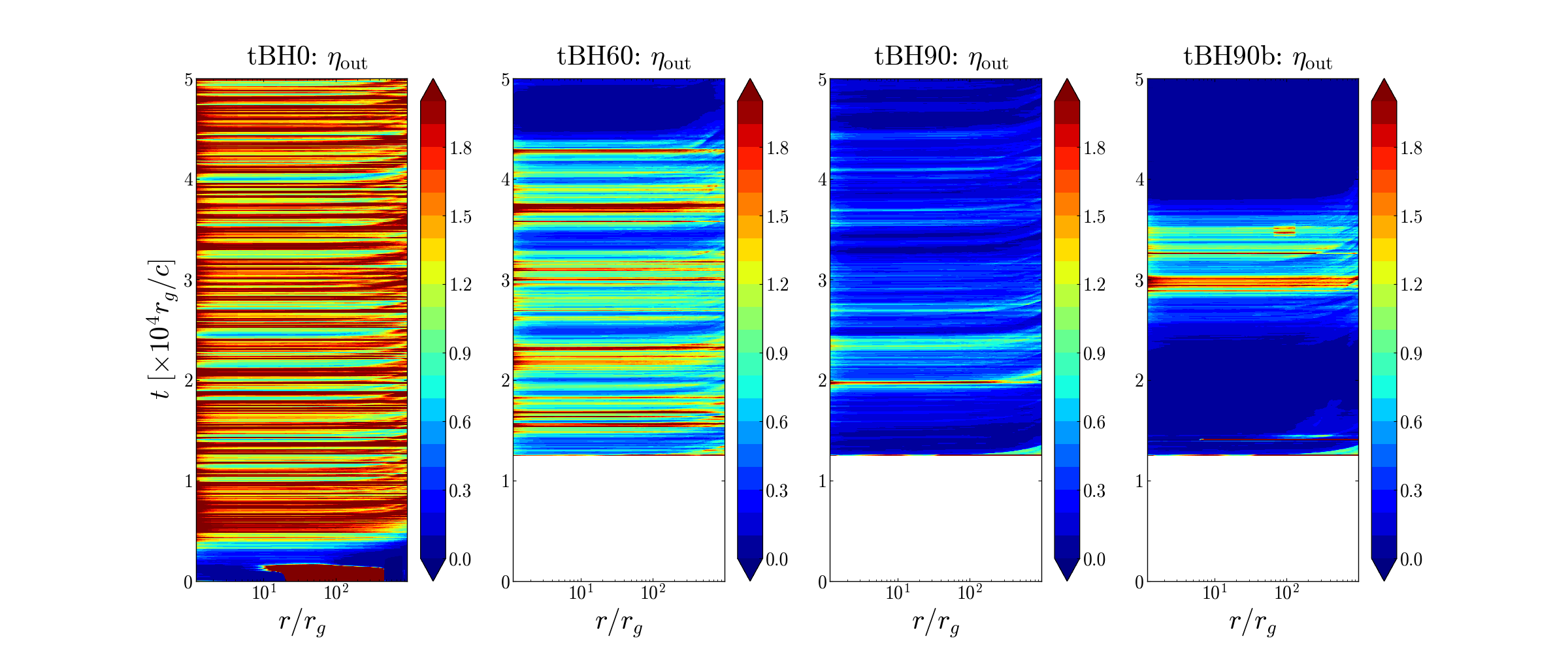}
    \includegraphics[width=\textwidth,trim= 0 0 0 0, clip]{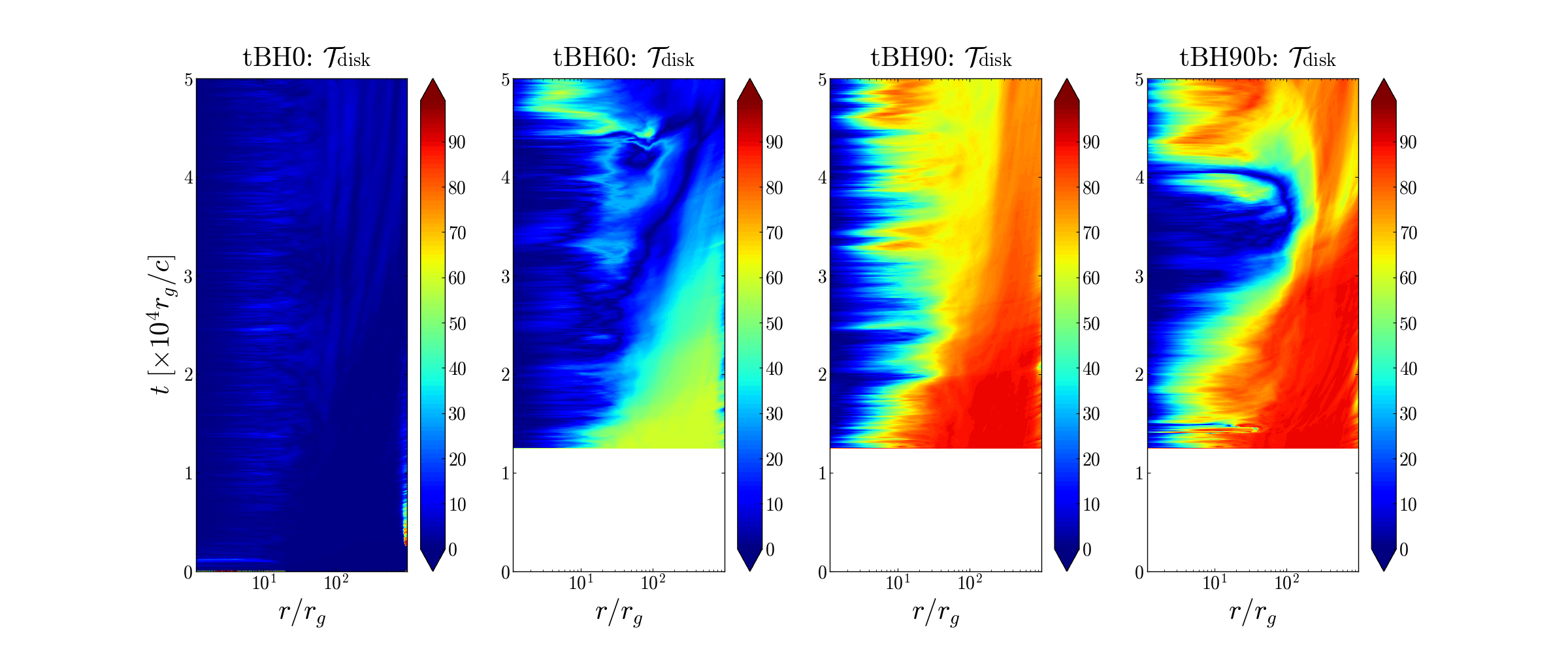}
    \caption{Space-time plots (similar to Fig.~\ref{fig:phibh}) of tilted black hole simulations, where instead of tilting the disk, we tilt the BH spin axis after the inner flow attains the MAD state.}
    \label{fig:tiltbh}
\end{figure*}

\section{Does misalignment depend on accretion history?}
\label{sec:accretion_history}

\begin{figure*}
    \centering
    \includegraphics[width=0.45\textwidth,trim= 0 0 0 0, clip]{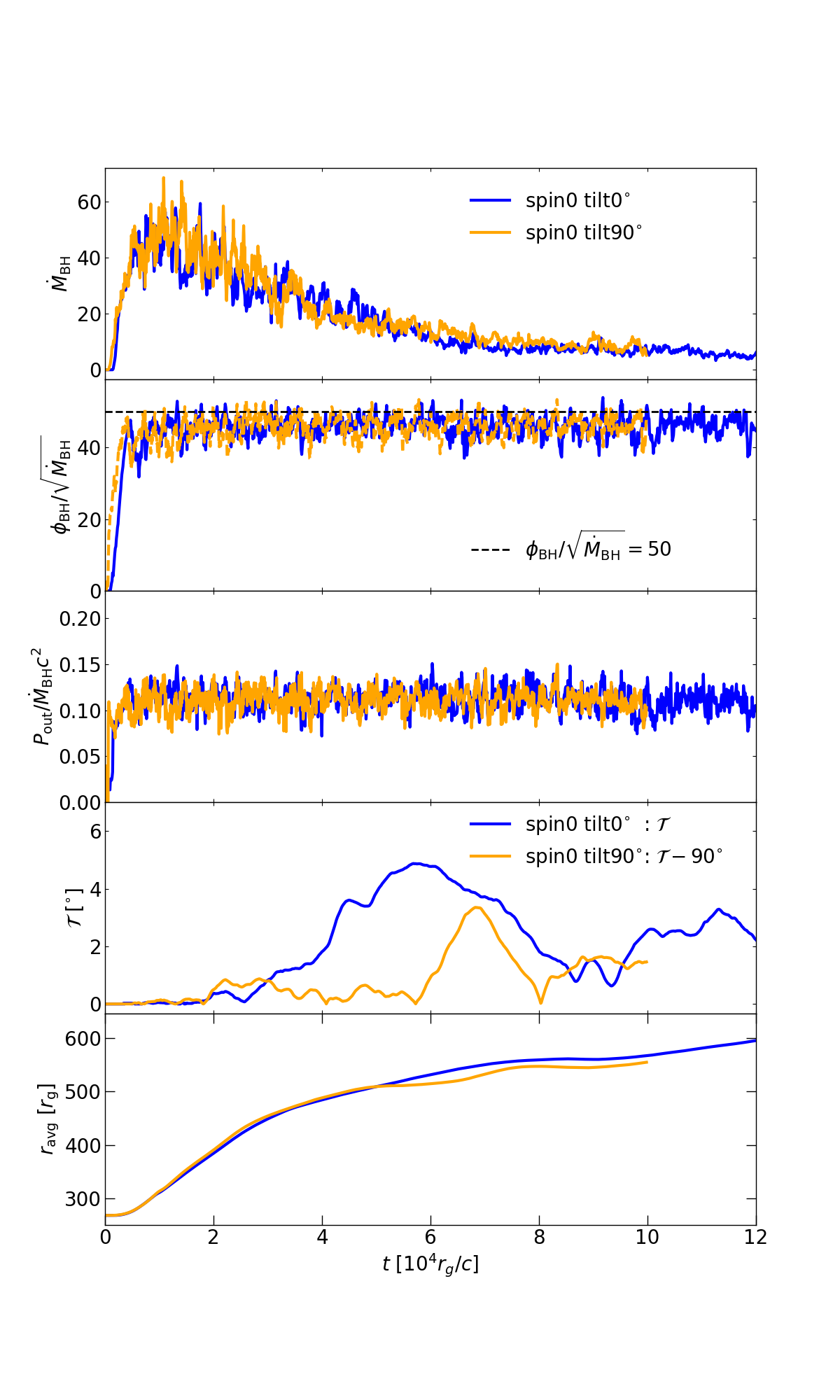}
    \includegraphics[width=0.45\textwidth,trim= 0 0 0 0, clip]{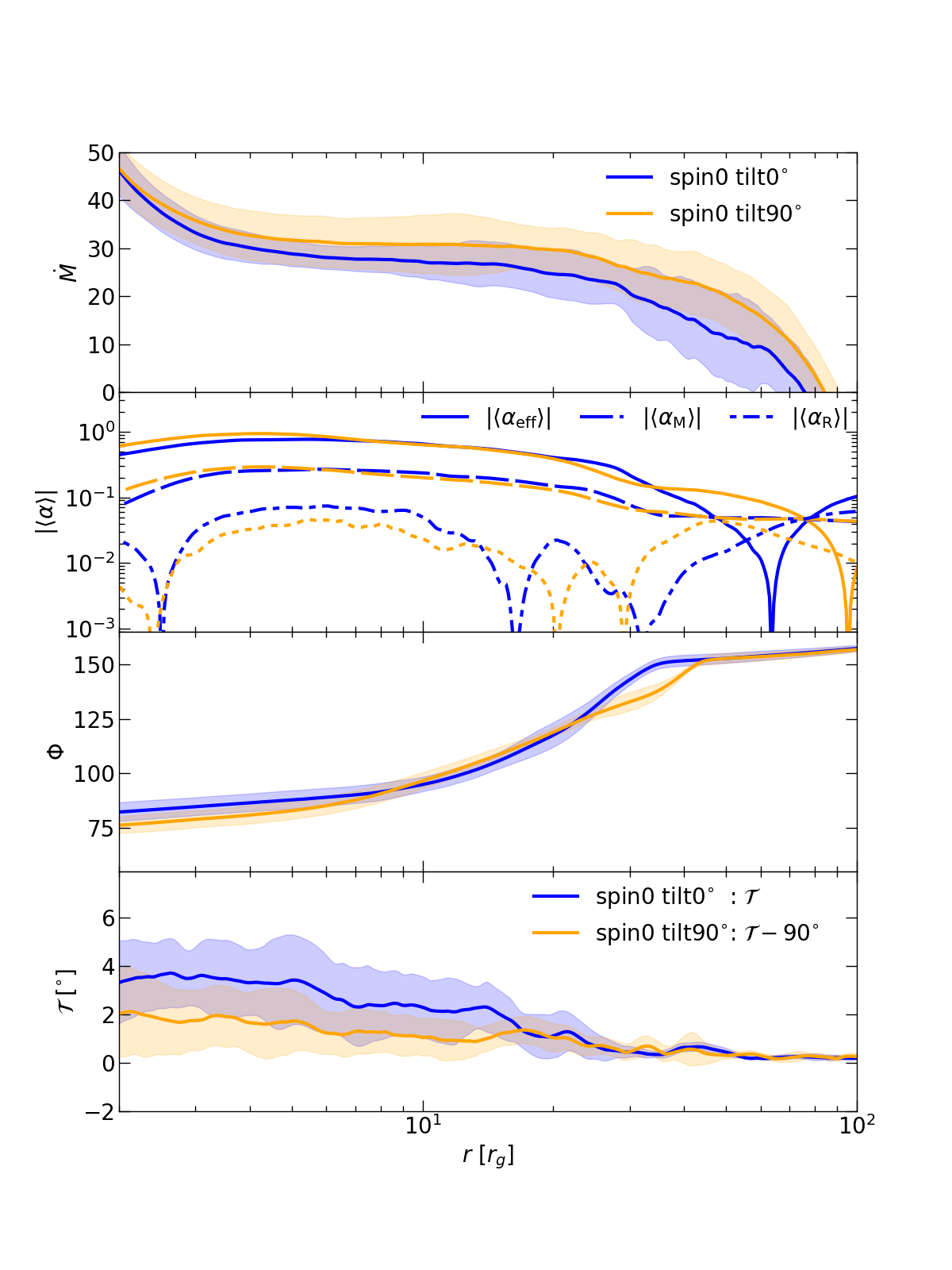}
    
    \caption{A comparison between a disk aligned to the polar grid midplane $\theta=\pi/2$ and a disk aligned with the polar axis $\theta=0$ for a Schwarzschild BH. We look at multiple time and radial plots of disk properties and find that our grid setup behaves well when the disk is tilted.}
    \label{fig:zero_spin}
\end{figure*}
In our fiducial model set, we initialize the BH with a spin vector along the $z$-axis and rotate the disk by an initial misalignment angle. To test the robustness of our results, we perform 3 moderate-resolution ($580\times 192\times 384$) simulations where we rotate the BH spin vector instead of the disk. We dub these models as \texttt{TxBH} where `\texttt{x}' ($\equiv$ 0, 60 and 90) is the BH misalignment angle with respect to the $z$-axis in degrees. Additionally we follow the approach of \citet{mckinney_2013} and only change the BH spin orientation \textit{after} the initially non-tilted BH-disk (model $\tiltBHa{}$) system reaches the MAD state at $t=2.5\times10^4\rg/c$. Such an approach will also test whether our results from the fiducial simulations is robust against accretion history, i.e., does alignment depend on whether the initial flow was in the magnetically arrested state? 

Figure~\ref{fig:tiltbh} shows that the tilted BH models behave similar to the ``tilted disk'' simulations over all: (i) the non-tilted $\tiltBHa{}$ continues to show powerful jets; (ii) the $60^{\circ}$ tilt model, $\tiltBHb{}$ considerably aligns with the BH midplane and displays prominent jets until $t\sim4.5\times10^4\rg/c$. As we see from the disk magnetic flux plot of $\tiltBHb{}$, around the same time as the jet ceases, there is a significant loss of magnetic flux in the disk, indicating a jet-disk collision; (iii) The $90^{\circ}$ model, $\tiltBHc{}$ does not really exhibit any prominent jets. However when we decreased the plasma$-\beta$ by a factor of 10 in model $\tiltBHd{}$, we see similar jet-disk collisions as seen in model $\tg{}$. While more rigorous studies are required, our results seem to indicate that accretion history does not matter for alignment when the magnetic flux supply is large. However, if the flow is marginally MAD, there is a possibility that the flow becomes non-MAD upon introducing misalignment.

\section{Grid verification: the Schwarzschild case}
\label{sec:zero_spin}

Here we check how well our grid performs as we tilt our disk from the equatorial midplane at $\theta=\pi/2$ to the pole at $\theta=0$ in the case of a spin-zero BH. Since the spin is zero, ideally these two disks should evolve similarly barring stochastic turbulence. Figure~\ref{fig:zero_spin} shows that all of our disk properties match remarkably well between the two disks, especially the radial profiles of the Maxwell, Reynolds and the total effective $\alpha$ viscosities, providing confidence that our grid setup resolves the turbulence sufficiently well, and the polar axis is properly transmissive. 

\bibliography{Refs-MAD_Tilted.bib}

\end{document}